\documentclass[twocolumn]{aastex631}
\usepackage{xfrac,graphicx,color,float,hyperref}
\usepackage{soul} 
\usepackage{lineno}
\usepackage{seqsplit}

\received{June 5, 2025}

\submitjournal{ApJ}

\begin{document}

\title{COSMOS-Web: Comprehensive Data Reduction for Wide-Area JWST NIRCam Imaging}

\shortauthors{Franco et al.}

\correspondingauthor{M. Franco}
\suppressAffiliations

\email{maximilien.franco@cea.fr}

\author[0000-0002-3560-8599]{Maximilien Franco}
\affiliation{Université Paris-Saclay, Université Paris Cité, CEA, CNRS, AIM, 91191 Gif-sur-Yvette, France}
\affiliation{The University of Texas at Austin, 2515 Speedway Blvd Stop C1400, Austin, TX 78712, USA}
\author[0000-0002-0930-6466]{Caitlin M. Casey}
\affiliation{Department of Physics, University of California, Santa Barbara, Santa Barbara, CA 93109, USA}
\affiliation{The University of Texas at Austin, 2515 Speedway Blvd Stop C1400, Austin, TX 78712, USA}
\affiliation{Cosmic Dawn Center (DAWN), Denmark}
\author[0000-0002-6610-2048]{Anton M. Koekemoer}
\affiliation{Space Telescope Science Institute, 3700 San Martin Dr., Baltimore, MD 21218, USA}
\author[0000-0001-9773-7479]{Daizhong Liu}
\affiliation{Purple Mountain Observatory, Chinese Academy of Sciences, 10 Yuanhua Road, Nanjing 210023, China}

\author[0000-0002-9921-9218]{Micaela B. Bagley}
\affiliation{The University of Texas at Austin, 2515 Speedway Blvd Stop C1400, Austin, TX 78712, USA}

\author[0000-0002-9489-7765]{Henry Joy McCracken}
\affiliation{Institut d’Astrophysique de Paris, UMR 7095, CNRS, and Sorbonne Université, 98 bis boulevard Arago, F-75014 Paris, France}

\author[0000-0001-9187-3605]{Jeyhan S. Kartaltepe}
\affiliation{Laboratory for Multiwavelength Astrophysics, School of Physics and Astronomy, Rochester Institute of Technology, 84 Lomb Memorial Drive, Rochester, NY 14623, USA}

\author[0000-0003-3596-8794]{Hollis B. Akins}
\affiliation{The University of Texas at Austin, 2515 Speedway Blvd Stop C1400, Austin, TX 78712, USA}

\author[0000-0002-7303-4397]{Olivier Ilbert}
\affiliation{Aix Marseille Univ, CNRS, CNES, LAM, Marseille, France }

\author[0000-0002-7087-0701]{Marko Shuntov}
\affiliation{Cosmic Dawn Center (DAWN), Denmark} 
\affiliation{Niels Bohr Institute, University of Copenhagen, Jagtvej 128, DK-2200, Copenhagen, Denmark}

\author[0000-0003-0129-2079]{Santosh Harish}
\affiliation{Laboratory for Multiwavelength Astrophysics, School of Physics and Astronomy, Rochester Institute of Technology, 84 Lomb Memorial Drive, Rochester, NY 14623, USA}

\author[0000-0002-4271-0364]{Brant E. Robertson}
\affiliation{Department of Astronomy and Astrophysics, University of California, Santa Cruz, 1156 High Street, Santa Cruz, CA 95064, USA}

\author[0000-0002-0569-5222]{Rafael C. Arango-Toro}
\affiliation{Aix Marseille Univ, CNRS, CNES, LAM, Marseille, France  }

\author[0000-0003-4569-2285]{Andrew J. Battisti}
\affiliation{International Centre for Radio Astronomy Research (ICRAR), University of Western Australia, M468, 35 Stirling Highway, Crawley, WA 6009, Australia}
\affiliation{Australian National University, Research School of Astronomy and Astrophysics, Canberra, ACT 2611, Australia}
\affiliation{ARC Centre of Excellence for All Sky Astrophysics in 3 Dimensions (ASTRO 3D), Australia}

\author[0000-0003-3691-937X]{Nima Chartab}
\affiliation{Caltech/IPAC, 1200 E. California Blvd., Pasadena, CA 91125, USA}

\author[0000-0003-4761-2197]{Nicole E. Drakos}
\affiliation{Department of Physics and Astronomy, University of Hawaii, Hilo, 200 W Kawili St, Hilo, HI 96720, USA}

\author[0000-0002-9382-9832]{Andreas L. Faisst}
\affiliation{Caltech/IPAC, MS 314-6, 1200 E. California Blvd. Pasadena, CA 91125, USA}

\author{Carter Flayhart}
\affiliation{Laboratory for Multiwavelength Astrophysics, School of Physics and Astronomy, Rochester Institute of Technology, 84 Lomb Memorial Drive, Rochester, NY 14623, USA}

\author[0000-0002-0236-919X]{Ghassem Gozaliasl}
\affiliation{Department of Computer Science, Aalto University, P.O. Box 15400, FI-00076 Espoo, Finland}
\affiliation{Department of Physics, University of, P.O. Box 64, FI-00014 Helsinki, Finland}

\author[0000-0002-3301-3321]{Michaela Hirschmann}
\affiliation{Institute of Physics, GalSpec, Ecole Polytechnique Federale de Lausanne, Observatoire de Sauverny, Chemin Pegasi 51, 1290 Versoix, Switzerland}
\affiliation{INAF, Astronomical Observatory of Trieste, Via Tiepolo 11, 34131 Trieste, Italy}

\author[0000-0002-6085-3780]{Richard Massey}
\affil{Department of Physics, Centre for Extragalactic Astronomy, Durham University, South Road, Durham DH1 3LE, UK}

\author[0000-0002-4485-8549]{Jason Rhodes}
\affiliation{Jet Propulsion Laboratory, California Institute of Technology, 4800 Oak Grove Drive, Pasadena, CA 91001, USA}

\author[0000-0002-0364-1159]{Zahra Sattari}
\affiliation{Department of Physics and Astronomy, University of California, Riverside, 900 University Ave, Riverside, CA 92521, USA}
\affiliation{The Observatories of the Carnegie Institution for Science, 813 Santa Barbara St., Pasadena, CA 91101, USA}

\author[0000-0001-8450-7885]{Diana Scognamiglio}
\affiliation{Argelander-Institut f\"ur Astronomie, Auf dem H\"ugel 71, D-53121 - Bonn, Germany}

\author[0000-0003-1614-196X]{John R. Weaver}
\affil{Department of Astronomy, University of Massachusetts, Amherst, MA 01003, USA}

\author[0000-0002-8434-880X]{Lilan Yang}
\affiliation{Laboratory for Multiwavelength Astrophysics, School of Physics and Astronomy, Rochester Institute of Technology, 84 Lomb Memorial Drive, Rochester, NY 14623, USA}

\author[0000-0002-7051-1100]{Jorge A. Zavala}
\affiliation{University of Massachusetts Amherst, 
710 North Pleasant Street, Amherst, MA 01003-9305, USA}

\author[0000-0002-8494-3123]{Edward M. Berman}
\affiliation{Department of Physics, Northeastern University, 360 Huntington Ave, Boston, MA}

\author[0000-0002-8008-9871]{Fabrizio Gentile}
\affiliation{CEA, Université Paris-Saclay, Université Paris Cité, CNRS, AIM, 91191, Gif-sur-Yvette, France}
\affiliation{INAF- Osservatorio di Astrofisica e Scienza dello Spazio, Via Gobetti 93/3, I-40129, Bologna, Italy}

\author[0000-0001-9885-4589]{Steven Gillman}
\affiliation{Cosmic Dawn Center (DAWN), Denmark}
\affiliation{DTU-Space, Technical University of Denmark, Elektrovej 327, DK-2800 Kgs. Lyngby, Denmark}

\author[0000-0002-7530-8857]{Arianna S. Long}
\affiliation{Department of Astronomy, The University of Washington, Seattle, WA 98195, USA}

\author[0000-0002-4872-2294]{Georgios Magdis}
\affiliation{Cosmic Dawn Center (DAWN), Denmark} 
\affiliation{DTU-Space, Technical University of Denmark, Elektrovej 327, 2800, Kgs. Lyngby, Denmark}
\affiliation{Niels Bohr Institute, University of Copenhagen, Jagtvej 128, DK-2200, Copenhagen, Denmark}

\author[0000-0002-9883-7460]{Jacqueline E. McCleary}
\affiliation{Department of Physics, Northeastern University, 360 Huntington Ave, Boston, MA}

\author[0000-0002-6149-8178]{Jed McKinney}
\altaffiliation{NASA Hubble Fellow}
\affiliation{Department of Astronomy, The University of Texas at Austin, 2515
Speedway Blvd Stop C1400, Austin, TX 78712, USA}

\author[0000-0001-5846-4404]{Bahram Mobasher}
\affiliation{Department of Physics and Astronomy, University of California, Riverside, 900 University Avenue, Riverside, CA 92521, USA}

\author[0000-0003-2397-0360]{Louise Paquereau} 
\affiliation{Institut d’Astrophysique de Paris, UMR 7095, CNRS, and Sorbonne Université, 98 bis boulevard Arago, F-75014 Paris, France}

\author[0000-0002-4410-5387]{Armin Rest}
\affiliation{Space Telescope Science Institute, Baltimore, MD 21218, USA}
\affiliation{Department of Physics and Astronomy, The Johns Hopkins University, Baltimore, MD 21218, USA}

\author[0000-0002-1233-9998]{David B. Sanders}
\affiliation{Institute for Astronomy, University of Hawaii, 2680 Woodlawn Drive, Honolulu, HI 96822, USA}

\author[0000-0003-3631-7176]{Sune Toft}
\affiliation{Cosmic Dawn Center (DAWN), Denmark} 
\affiliation{Niels Bohr Institute, University of Copenhagen, Jagtvej 128, DK-2200, Copenhagen, Denmark}

\author[0000-0002-3462-4175]{Si-Yue Yu}
\affiliation{Kavli Institute for the Physics and Mathematics of the Universe (WPI), The University of Tokyo, Kashiwa, Chiba 277-8583, Japan}
\affiliation{Department of Astronomy, School of Science, The University of Tokyo, 7-3-1 Hongo, Bunkyo, Tokyo 113-0033, Japan}

\collaboration{40}{\vspace{-20pt}}


\begin{abstract} 
We present the data reduction methodology used for the COSMOS-Web survey JWST NIRCam data. Covering 0.54 deg$^2$ with four broadband filters (F115W, F150W, F277W, F444W) and a total exposure time of approximately 270 hours, COSMOS-Web represents the largest contiguous field surveyed during JWST Cycle 1, posing unique data reduction challenges due to its extensive scale. By combining the official JWST Calibration Pipeline with custom improvements for noise removal, background subtraction, and astrometric alignment, we achieve high fidelity science-ready mosaics. We detail the systematic approach employed in the three stages of the JWST Calibration Pipeline. The data, collected in three epochs from January 2023 to January 2024, encompass 152 visits and have been processed into 20 mosaic tiles to optimize computational efficiency and data processing.  The final data products achieve 5$\sigma$ depths of 26.7–28.3 AB mag in 0.15" apertures. The processed and calibrated datasets are made available to the public.
\end{abstract}


\section{Introduction}
\label{sec:intro}

Advancements in our understanding of galaxy formation and evolution have significantly accelerated with the advent of deep, multi-wavelength surveys. Among these, the Cosmic Evolution Survey (COSMOS; \citealt{Scoville2007, Koekemoer2007}) has been instrumental due to its combination of large area ($\sim$2 deg$^2$) and imaging depth, providing a statistically significant sample for probing galaxy evolution and cosmic structure formation from the local universe to high redshift. The COSMOS-Web treasury program (\citealt{Casey2023}, PIs: Kartaltepe \& Casey, ID=1727) with the James Webb Space Telescope (JWST) marks a significant milestone in this ongoing quest, aiming to provide deeper insights into the early universe and the formation of galaxies across cosmic time through the Near-Infrared Camera (NIRCam; \citealt{Rieke2003, Rieke2005, Beichman2012, Rieke2023}) over 0.54 deg$^2$ in four filters (F115W, F150W, F277W, F444W) and through the Mid-Infrared Instrument (MIRI; \citealt{Rieke2015, Wright2015}) over 0.20 deg$^2$ in a single broadband filter, F770W. 

Building on the legacy of COSMOS and its rich multi-wavelength coverage \citep{Scoville2007, Koekemoer2007, Ilbert2009, McCracken2012}, COSMOS-Web leverages the capabilities of JWST to probe earlier epochs and fainter galaxies than previously possible in the COSMOS field. The strategic focus on the COSMOS field offers not only depth and area, but also benefits from an extensive range of complementary data, from X-rays to radio, facilitating precise photometric redshifts via spectral energy distribution (SED) fitting and enabling the derivation of key physical parameters for statistically robust galaxy samples \citep{Capak2007, Ilbert2009, Laigle2016, McCracken2012, Muzzin2013, Weaver2022}. In addition, its equatorial position on the sky ensures optimal follow-up opportunities. It facilitates the integration of new survey data with existing multi-wavelength datasets, thereby enriching our multi-dimensional view of the universe \citep[e.g.,][]{Capak2007}.

The first few years of JWST observations have catalyzed a surge in new scientific breakthroughs, exploring a wide range of topics, from exoplanetary systems to the most distant galaxies observed to date \citep[e.g.,][]{Curtis2023, castellano22a, naidu22, whitler23, finkelstein22b,adams23, leung23, Casey2024, Carniani2024, Castellano2024, Stark2025, Napolitano2025}. These studies highlight just how powerful NIRCam data can be for exploring the early universe, especially when identifying galaxy candidates at $z \geq 10$. By pushing observations into these early epochs, we are opening up new opportunities to better understand the first 500\,Myr of the universe.

The task of data reduction, in this context, is more than a routine technical step. It is fundamental to ensuring the quality and scientific utility of the observations. The process involves a series of quality control steps designed to eliminate instrumental noise, correct for cosmic rays, artifacts, and improve the astrometric accuracy of the dataset. This is particularly challenging given the size of the dataset (12,880 individual raw files) and is essential in order to produce accurate photometric and morphological analyses \citep[e.g.,][]{Casey2024, Mercier2024, Akins2024, Lambrides2024, Arango2024, Silverman2023, Gentile2024, Faisst2025, Shuntov2025, Paquereau2025, Nightingale2025, Huertas-Company2025}. Furthermore, the need for rigorous data reduction is accentuated by the fact that COSMOS (and therefore COSMOS-Web) has been adopted as a standard calibration field for ongoing and future large surveys (e.g., \textit{Euclid}, Roman).

In this paper, we detail the procedures and techniques implemented for the reduction of NIRCam observations from the COSMOS-Web survey. A companion paper will describe the techniques used for MIRI data reduction (Harish et al., in prep.). It is structured to provide an account of the observational strategy, data reduction pipeline enhancements, and the scientific outcomes resulting from this work. Section~\ref{sec:observations} describes the observational strategy, detailing the survey design and the execution of observations. Section~\ref{sec:data_reduction} explores the process of data reduction, including the challenges faced and the solutions implemented. Finally, Section~\ref{sec:results} presents the results.

\begin{figure*}[!]
\centering
\includegraphics[width=0.9\textwidth]{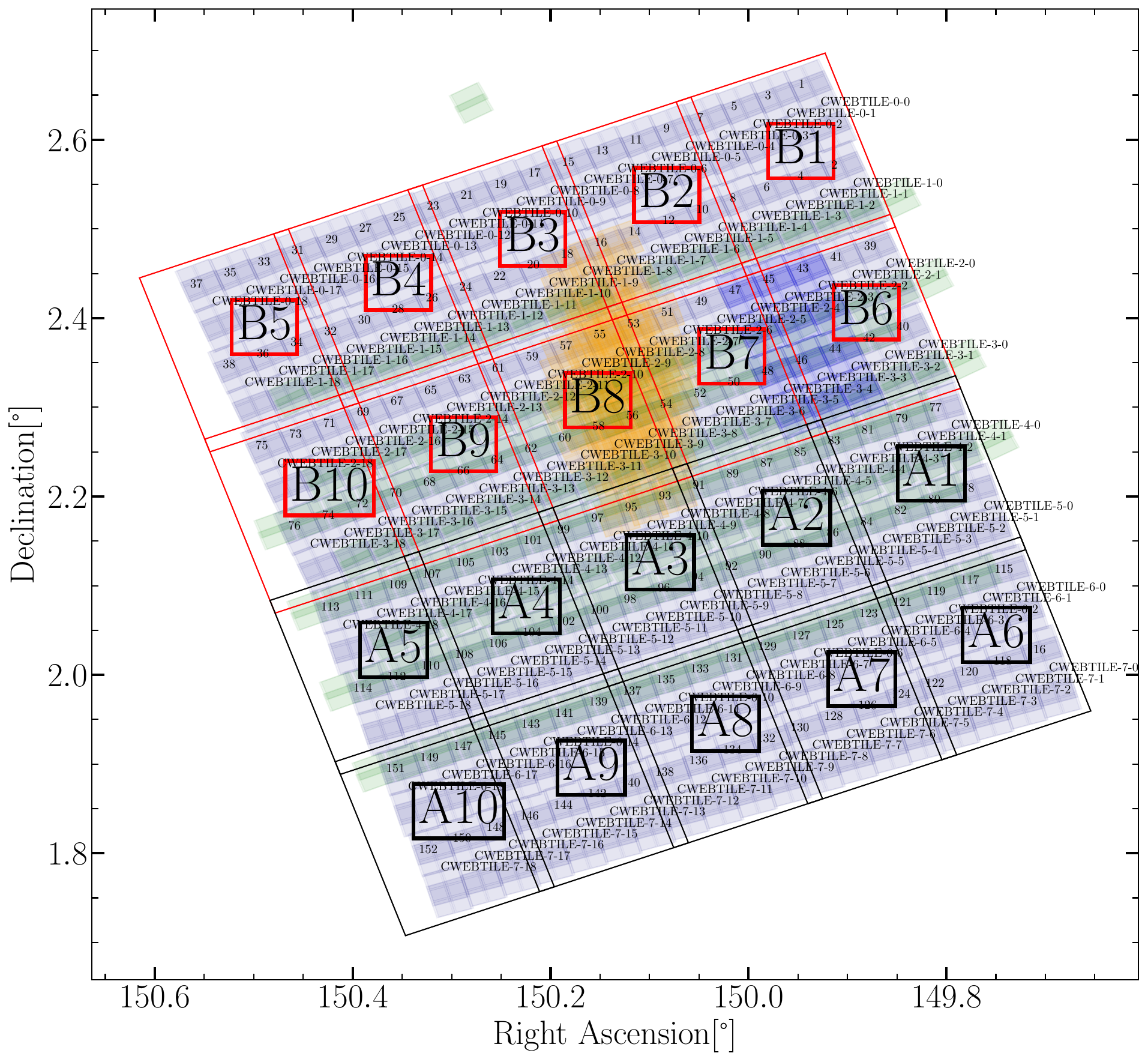} 
\caption{Exposure map presenting the COSMOS-Web NIRCam observation pattern design, depicting NIRCam SW filter observations in blue with MIRI parallels in green. NIRCam LW coverage is largely identical to SW,  although slightly simpler by virtue of larger detectors.  Observation tiles for April 2023 and January 2024 are marked in black and red, respectively, while the initial dataset from January 2023 is highlighted in darker blue.  In addition, we overlay the NIRCam PRIMER \citep{Dunlop2021} positions in yellow. The name and the number of each visit are indicated for clarity and reference, also provided in the appendix of \citet{Casey2023} as a table.}
         \label{Fig::CW_observations}
\end{figure*}

\begin{figure*}[!]
\centering
\includegraphics[width=1\textwidth]{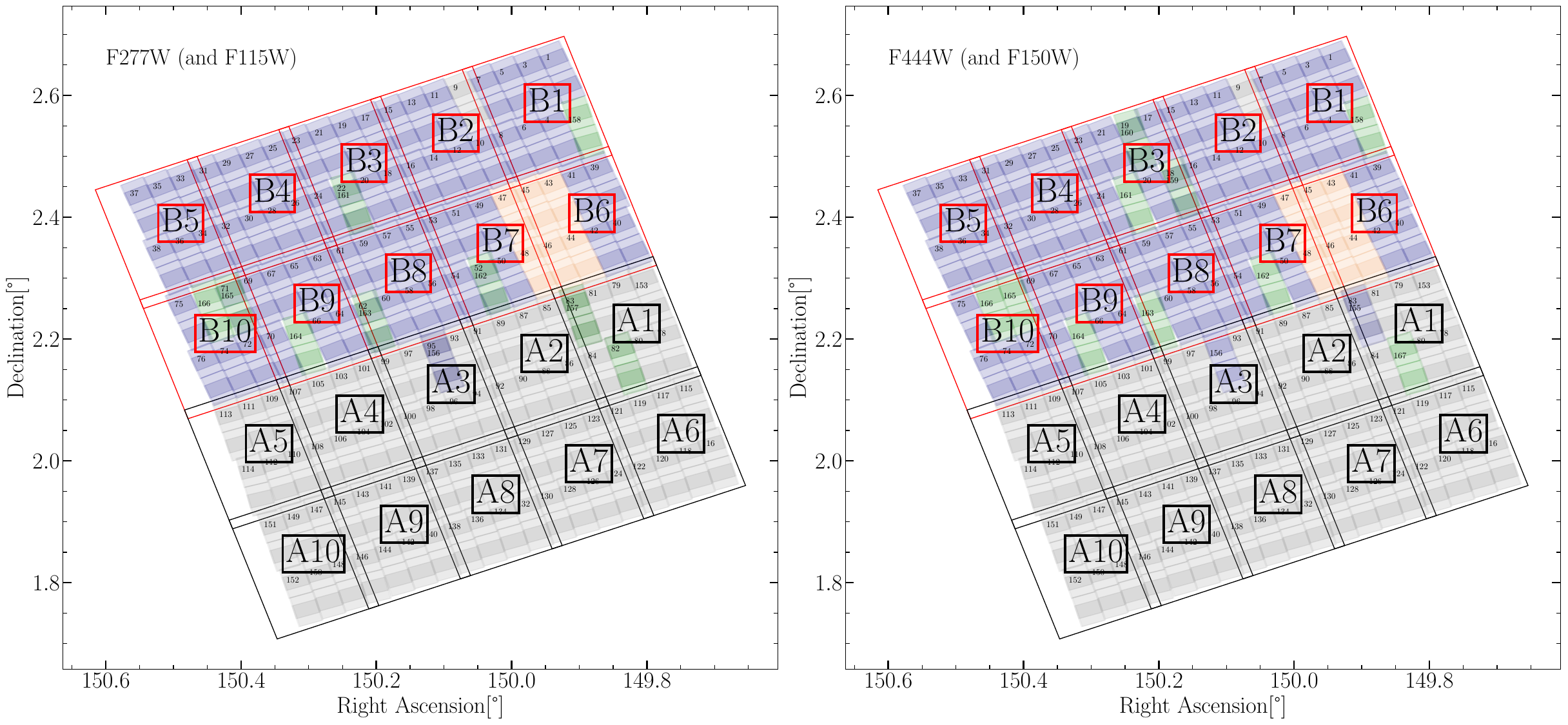} 
\caption{Effective coverage achieved by the COSMOS-Web survey using the NIRCam filters, illustrating the spatial distribution of observations. In the left panel, the coverage in the F277W filter is displayed, which also corresponds to the coverage achieved with the F115W filter. The right panel shows the survey's extent in the F444W filter, which aligns with the coverage in the F150W filter. The first epoch of observations, from January 5, 2023, to January 6, 2023, is indicated in orange. The second period, from April 6 to April 23, 2023 (with an additional visit, CWEBTILE-0-4, on May 17, 2023), is shown in gray. The third epoch spans from December 12 2023 to January 7 2024 and is displayed in blue. The visits that were missed and observed afterward from April 5 2024 to May 17 2024 (see Section~\ref{sec::Failed_Observations}) are marked in green. When a visit formally failed and required a Webb Operation Problem Reports (WOPR) approval to be re-observed, it was given a new visit number; in these cases, the numbers of both visits are indicated one above the other. Additional elements depicted in this figure are consistent with those outlined in Fig.~\ref{Fig::CW_observations}.}
         \label{Fig::Effective_coverage}
\end{figure*}

\section{Observations}
\label{sec:observations}

\subsection{COSMOS-Web design}

COSMOS-Web's survey design incorporates four NIRCam filters (F115W, F150W, F277W, and F444W) to image the largest contiguous area to date with JWST, setting it apart from other extragalactic pencil-beam surveys like PRIMER (GO\#1837; \citealt{Dunlop2021}), JADES (GTO \#1180, 1181, 1210, 1286, 1287; \citealt{Eisenstein2023}), and CEERS (ERS \#1345 \citealt{Finkelstein2025}) over smaller areas. The choice of using four filters is a balance between depth, spectrum sampling, and coverage area, allowing the survey to span a contiguous mosaic of 0.54 deg$^2$ through the NIRCam filters. Additionally, the survey is complemented by MIRI imaging, with one filter (F770W), extending over a noncontiguous 0.20 deg$^2$ area, which will be discussed in a separate paper (Harish et al., in prep.).

The NIRCam observations are designed as a contiguous 41.5' $\times$ 46.6' rectangular mosaic 
centered at RA = 10:00:27.92, Dec = +02:12:03.5 oriented with a positional angle of 20 degrees. The mosaic covers 152 visits arranged into 19 columns and 8 rows (see Fig.~\ref{Fig::CW_observations}). Further details on the survey design and its scientific goals can be found in the survey overview paper \citep{Casey2023}.

Each visit is organized into two sequences using the 4-TIGHT dither pattern (closely spaced positions to improve spatial sampling and mitigate detector artifacts), with each sequence comprising four integrations; each lasting approximately 257 seconds. The NIRCam filters are paired as F115W (SW) with F277W (LW) and F150W (SW) with F444W (LW) for observations. This configuration implies that for the short wavelength (SW) filters, each visit has 32 individual exposures (across 8 detectors and 4 dithers), whereas the long wavelength (LW) filters involve 8 exposures (across 2 detectors and 4 dithers). Consequently, the survey accumulates a total of 80 exposures per visit, totaling 12,880 exposures over the entire survey. This effective number is slightly higher than the expected 12,160 exposures, as it includes additional visits that were made to compensate for those interrupted.

The implementation of the 4-TIGHT dither pattern, aimed at maximizing the contiguous coverage area, results in non-uniform observational coverage across the survey field, as illustrated by the varying shades of blue in Fig.~\ref{Fig::CW_observations}. For the LW (SW) filters, 51\% (51\%) of the survey area is covered by 2 exposures, 47\% (42\%) is covered by 4 exposures. The area covered by odd numbers of exposures (1 or 3 exposures) is significantly smaller with 2\% (7\%) of the area of the survey \citep{Casey2023}.

\subsection{Observation Scheduling}

The COSMOS-Web survey was planned for 255 hours of observations but required 270.3 hours to be completed  due to differences in actual overheads required versus those originally assumed when planning the program.  The COSMOS field is accessible to JWST observations in only two windows during the year: one spanning most of April and May and the other spanning late November until early January. During these two epochs, there are fairly restrictive instrument PAs that are roughly 180° apart.  The COSMOS-Web observations were designed to be carried out during the lowest-background part of these windows; therefore most observations were either carried out in early April or early January.

The first epoch of COSMOS-Web, designated as Jan23, is comprised of six visits (visits numbered 43 to 48 in the Astronomer's Proposal Tool - APT) that surveyed an area of approximately 77 arcmin$^2$. These initial observations were conducted on January 5–6, 2023 and led to some early scientific analyses \citep{Silverman2023,McKinney2023,Akins2023, Franco2024}. During the second epoch, Apr23, a further 77 pointings were carried out (visits numbered 77 to 152 and visit 9 in the APT), covering roughly half of the final survey. This phase took place from April 7th to May 17th 2023. The final epoch, referred to as Jan24, included the remaining 69 pointings (visits numbered 1-8, 10-42 then 49-76), extending from December 12 2023 to January 7 2024.  The visits that were missed were observed afterwards from April 5 2024 to May 17 2024 (see Section~\ref{sec::Failed_Observations}), thereby marking the completion of the observational campaign.

For clarity throughout this document, the epochs will be referred to by their respective abbreviations: Jan23 for the initial phase in January 2023, Apr23 for the second phase in April-May 2023, and Jan24 for the final phase in December 2023-January 2024.
All failed visits completed after the Jan24 epoch have been fully folded in to the final mosaics and analysis here and treated as part of the designated epoch they were originally planned for.

\subsection{Failed Observations}
\label{sec::Failed_Observations}
Unfortunately, not all visits were successful on the first attempt. The initial series of observations conducted in January 2023 were completed as planned. Three of the Apr23 visits failed and seven of the Jan24 visits failed because of guiding star problems Table~\ref{tab::failed_visits} summarizes the failed visits, their new IDs in the completed program, and some brief explanation of the particular issues or reasons the initial observations were rejected.

\begin{deluxetable*}{ccl}
 \tabletypesize{\scriptsize}
 \tablecaption{Failed visits \label{tab::failed_visits}}
 \tablehead{
 \colhead{Original ID} & \colhead{Final ID} & \colhead{Usable Data Details}
 }
\startdata
2&158 & No NIRCam exposures obtained. \\
18&159 & All F115W+F277W obtained; half of F150W+F444W.\\
19&160 & All F115W+F277W obtained; half of F150W+F444W.\\
22&161 & A quarter of F115W+F277W obtained; no F150W+F444W.\\
52&162 & Half of F115W+F277W obtained; no F150W+F444W.\\
62&163 & Half of F115W+F277W obtained; no F150W+F444W.\\
68&164 & No NIRCam exposures obtained.\\
71&165 & Half of F115W+F277W obtained; no F150W+F444W.\\
73&166 & No NIRCam exposures obtained.\\
77&153 & No NIRCam exposures obtained. \\
82&167 & Half of F115W+F277W obtained; no F150W+F444W. \\
83&155 & Half of F115W+F277W obtained; no F150W+F444W. \\
83&157 & All F115W+F277W obtained; half of F150W+F444W. \\
95&156 & All of F115W+F277W obtained; no F150W+F444W.\\
\enddata
\tablecomments{New designations for the visits that were missed or partially observed, and subsequently re-observed. Visit 83 was reobserved in two parts (see Fig.~\ref{Fig::Effective_coverage}).}
\end{deluxetable*}

Note that repeated visits were often taken during a different observing window, therefore requiring a flip in the PA to suit the alternate time of year.  As discussed in \citet{Casey2023}, the design of the mosaic is such that the Apr23 epoch observations are 6 degrees offset from the Jan23 epoch observations (both are 3° offset from the mosaic average PA of 20°).  In most cases, failed visits are repeated with a 180° flip from the failed visit, but in the final epoch of observations completed after Jan24, PAs were mistakenly not flipped by the full 180°; this results in a few final 'gaps' in the mosaic shaped like isosceles triangles measuring at most $\sim$128$''$ tall and $\sim$13$''$ along the base.

\subsection{Tiles}
\label{subsec::tile}

The COSMOS-Web survey, characterized by its large data volume, required a customized approach to data management and processing. Given the impracticality of creating and operating with a single large mosaic for most applications, due to computational efficiency constraints, the survey field was segmented into smaller tiles of manageable size, such that typical users and laptops may easily be able to load and manipulate data. In addition, this strategy enhances data processing efficiency and optimizes computational resource allocation.

The survey area was divided into 20 tiles, labeled A1 to A10 and B1 to B10. Tiles A1 - A10, displayed in black in Fig.~\ref{Fig::CW_observations} and Fig.~\ref{Fig::Effective_coverage}, cover the region observed in the Apr23 epoch in the southern half of COSMOS-Web. Conversely, tiles B1 - B10, displayed in red in Fig.~\ref{Fig::CW_observations}, span the areas observed in Jan23 and Jan24 epochs (the northern half of the field). Each of the 20 tiles measures 20.2 by 26.2 arcmin (approximately 530 arcmin$^2$ including overlap)  and is oriented at a 20-degree angle to maximally align with the survey layout and minimize the number of tiles required.

To mitigate edge effects, each tile is designed to overlap with its adjacent tiles by approximately 2 arcmin. This overlap facilitates data processing and analysis across the boundaries of the tiles (in order to avoid galaxies split in 2 different tiles). Detailed coordinates for each tile's corners are provided in Table~\ref{tab::tile_coordinates}, offering a framework for the survey's geometric configuration and serving as a potential reference for future datasets adopting the same tiling scheme. The full tiling layout is also available as a DS9 region file on the COSMOS-Web data release 1 (DR1) page.

\begin{deluxetable*}{lcccc}
 \tabletypesize{\scriptsize}
 \tablecaption{Tile coordinates \label{tab::tile_coordinates}}
 \tablehead{
 \colhead{ID} & \colhead{Corner 1 (RA, Dec)} & \colhead{Corner 2 (RA, Dec)} & \colhead{Corner 3 (RA, Dec)} & \colhead{Corner 4} (RA, Dec)
 }
\startdata
A1  & 149.8703317, 2.0856512 & 149.7198796, 2.1403395 & 149.7908786, 2.3354095 & 149.9413496, 2.2807163 \\
A2  & 150.0058959, 2.0363591 & 149.8554506, 2.0910612 & 149.9264667, 2.2861269 & 150.0769300, 2.2314186 \\
A3  & 150.1414523, 1.9870553 & 149.9910155, 2.0417704 & 150.0620479, 2.2368306 & 150.2125019, 2.1821081 \\
A4  & 150.2769995, 1.9377408 & 150.1265729, 1.9924679 & 150.1976208, 2.1875215 & 150.3480637, 2.1327859 \\
A5  & 150.4125359, 1.8884166 & 150.2621212, 1.9431545 & 150.3331838, 2.1382005 & 150.4836139, 2.0834528 \\
A6  & 149.8045087, 1.9048087 & 149.6540746, 1.9594923 & 149.7250552, 2.1545612 & 149.8755087, 2.0998725 \\
A7  & 149.9400575, 1.8555218 & 149.7896293, 1.9102182 & 149.8606274, 2.1052826 & 150.0110740, 2.0505800 \\
A8  & 150.0755992, 1.8062243 & 149.9251788, 1.8609325 & 149.9961935, 2.0559913 & 150.1466316, 2.0012757 \\
A9  & 150.2111325, 1.7569171 & 150.0607214, 1.8116361 & 150.1317520, 2.0066883 & 150.2821799, 1.9519607 \\
A10 & 150.3466557, 1.7076011 & 150.1962556, 1.7623299 & 150.2673014, 1.9573744 & 150.4177173, 1.9026358 \\
B1  & 150.0020274, 2.4473359 & 149.8515406, 2.5020333 & 149.9225757, 2.6970916 & 150.0730806, 2.6423895 \\
B2  & 150.1376214, 2.3980335 & 149.9871430, 2.4527469 & 150.0581944, 2.6478011 & 150.2086900, 2.5930817 \\
B3  & 150.2732061, 2.3487174 & 150.1227378, 2.4034461 & 150.1938048, 2.5984949 & 150.3442894, 2.5437590 \\
B4  & 150.4087801, 2.2993886 & 150.2583236, 2.3541315 & 150.3294054, 2.5491739 & 150.4798772, 2.4944226 \\
B5  & 150.5443418, 2.2500480 & 150.3938989, 2.3048040 & 150.4649946, 2.4998389 & 150.6154520, 2.4450733 \\
B6  & 149.9361713, 2.2664951 & 149.7857017, 2.3211879 & 149.8567188, 2.5162544 & 150.0072070, 2.4615567 \\
B7  & 150.0717506, 2.2171978 & 149.9212885, 2.2719056 & 149.9923224, 2.4669678 & 150.1428020, 2.4122539 \\
B8  & 150.2073213, 2.1678878 & 150.0568686, 2.2226097 & 150.1279183, 2.4176665 & 150.2783878, 2.3629373 \\
B9  & 150.3428821, 2.1185662 & 150.1924404, 2.1733011 & 150.2635052, 2.3683514 & 150.4139629, 2.3136080 \\
B10 & 150.4784314, 2.0692337 & 150.3280023, 2.1239807 & 150.3990815, 2.3190234 & 150.5495255, 2.2642668 \\
\enddata
\tablecomments{RA and Dec coordinates for the 20 Tiles Segmenting the COSMOS-Web Survey. Tiles A1 to A10, indicated in black in Fig.~\ref{Fig::CW_observations}, correspond to the region observed in April 2023. Tiles B1 to B10, highlighted in red in Fig.~\ref{Fig::CW_observations}, cover Jan23 and Jan24 areas.}
\end{deluxetable*}

\section{Data Reduction}
\label{sec:data_reduction}
The volume of data from COSMOS-Web introduces unique challenges in processing and analysis. 
The raw data were processed using the JWST Calibration Pipeline\footnote{\url{https://github.com/spacetelescope/jwst}} \citep{Bushouse2023}, supplemented with some custom steps and techniques inspired (and/or developed) by \cite{Bagley2023}. This section elaborates on these methodologies and their impact on enhancing data quality.

\subsection{JWST Pipeline and Calibration Reference Data System (CRDS) Implementation}

The processing of new observational epochs from the COSMOS-Web survey followed an incremental methodology, employing the latest version of the JWST pipeline and the Calibration Reference Data System (CRDS)\footnote{\url{https://jwst-crds.stsci.edu/}} at the time of the observations. 

For the initial observational epoch (Jan23), data processing was conducted using version 1.8.3 of the JWST pipeline \citep{Bushouse2023}, released by the Space Telescope Science Institute (STScI). We used the CRDS pmap 1017, corresponding to the NIRCam instrument mapping (imap) 0233. Subsequent data (Apr23) were processed with an updated version of the JWST pipeline, version 1.10.0, alongside CRDS version 1075 (imap 0252). For data obtained in Jan24, we used the JWST pipeline 1.12.1  alongside CRDS version 1170 (imap 0273). For the final processed images presented (the COSMOS-Web release DR1), we use JWST pipeline 1.14.0 alongside CRDS version 1223 (imap 0285).

\subsection{Pipeline Level 1}
Uncalibrated NIRCam raw data for the survey visits are retrieved from the Mikulski Archive for Space Telescopes (MAST). The stage 1 of the JWST pipeline performs detector-level corrections to produce a count-rate image in units of counts per second from uncalibrated images (raw ramps for all integrations). The different steps of this level include Data quality initialization, Saturation check, Reference pixel correction, Linearity correction, Jump detection and Slope fitting. 

In addition to the standard pipeline procedures, we have integrated a step for the subtraction of ‘snowballs' (Section~\ref{subsec::Snowball}) and ‘wisps', as detailed in Section~\ref{subsec::wisps},  the removal of 1/f noise, described in Section~\ref{subsec::1_f}, as well as a further step to remove the persistence (Section~\ref{subsec::persistence}) and  the correction of ‘claw’ artifacts (Section~\ref{subsec::claws}).

\subsubsection{Correction of Snowball Events in NIRCam Data}
\label{subsec::Snowball}
Snowball events in NIRCam data appear as extensive, roughly circular regions of elevated signal caused by high-energy cosmic-ray strikes that deposit charge across up to several hundred pixels \citep{Regan2023}. These phenomena present a challenge in the data processing pipeline due to their size and shape.

To mitigate the impact of snowball events, we explored optimal parameters for the Jump Detection step within the JWST Calibration Pipeline. Key adjustments made to the default parameters included:
\begin{itemize}\setlength\itemsep{-0.5em}
\item \texttt{expand\_factor = 2.2}
\item \texttt{max\_jump\_to\_flag\_neighbors = 300}
\item \texttt{min\_jump\_to\_flag\_neighbors = 15}
\item \texttt{min\_jump\_area = 15}
\item \texttt{sat\_required\_snowball = false}
\item \texttt{expand\_large\_events = true}
\item \texttt{\seqsplit{min\_sat\_radius\_extend = 2}}
\end{itemize}

Despite these adjustments, not all snowball events were effectively removed by this step. In some cases, snowballs may have occurred during the final readout, making them difficult to detect, or extended across more readouts than anticipated, complicating their identification and masking. Consequently, a manual inspection was implemented during the mosaic creation phase. For this purpose, we employed the \texttt{ShowerMasking} tool\footnote{\url{https://github.com/STScI-MIRI/ShowerMasking}}, which is specifically designed to facilitate the masking of cosmic ray showers. This tool proves invaluable in identifying and masking cosmic ray impacts that elude detection or removal by the standard JWST Calibration Pipeline, thereby enhancing the overall quality of the NIRCam data.

\begin{figure*}[!]
\centering
\includegraphics[width=1.\textwidth]{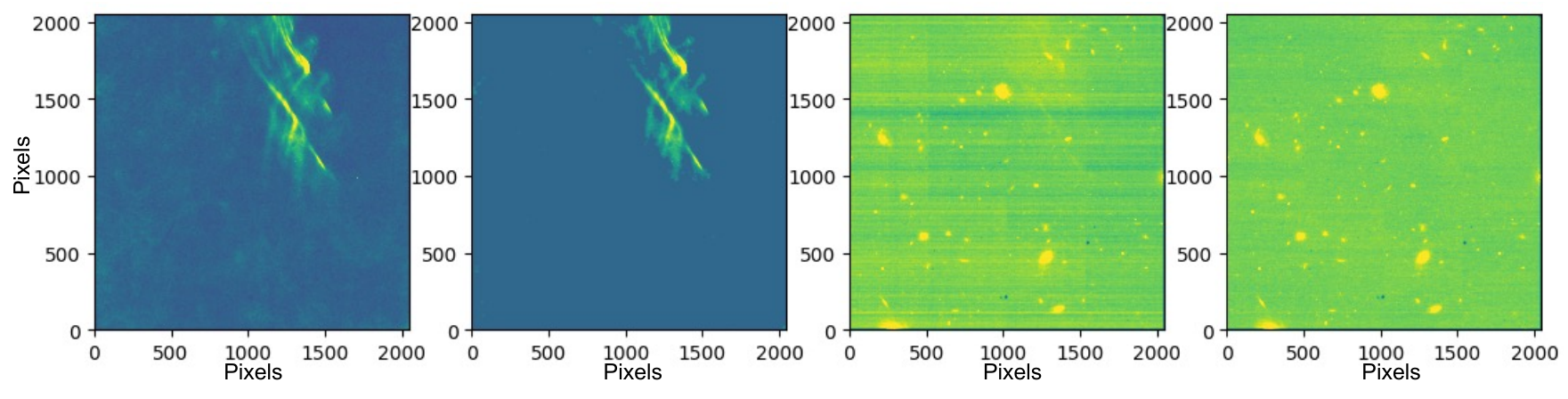} 
\caption{Illustration of the wisp correction process for the B4 detector of the F150W filter. The leftmost image displays the original wisp template for the detector. The  second from the left shows the wisp template after background removal, as detailed in Section~\ref{subsec::wisps}. The middle-right image represents the rate image prior to the correction of 1/f noise and wisps. Finally, the rightmost image demonstrates the rate image following the application of corrections for both 1/f noise and wisps.}         \label{Fig::Wisps}
\end{figure*}

\subsubsection{Correction of `Wisps'}
\label{subsec::wisps}

`Wisps' manifest as elongated, cirrus-like features \citep{Robotham2023} across the field of view in NIRCam images \citep{Robotham2023}\footnote{\url{https://jwst-docs.stsci.edu/known-issues-with-jwst-data/nircam-known-issues/nircam-scattered-light-artifacts}}, particularly near the edges. These artifacts arise from secondary scattered light, specifically from reflections off the structures supporting the secondary mirror \citep{Rigby2023}. Unlike other forms of scattered light that depend on the presence of bright astronomical sources, `wisps' are not directly linked to bright objects in the field. As a result, their shape and position are fixed in the detector plane, although their intensity can vary between exposures. `Wisps' are a recurring feature in the B4 detectors, with especially pronounced visibility in the F150W filter of the COSMOS-Web survey. Their presence and variability are less consistent across other detectors in the SW filters and are notably absent in the LW detectors \citep{Rigby2023, Robotham2023}.

The presence of wisps can significantly affect data analysis by artificially elevating the background level and introducing uncertainties in photometric measurements. To address this issue, we built wisp templates by median stacking all the COSMOS-Web observations available for each detector and each SW filter after correcting for 1/f noise.  We kept only the templates for which the presence of wisps is most pronounced and statistically significant (in particular, this applied to A1, A3, A4, B3 and B4 for F115W and F150W). These templates, derived from stacking JWST NIRCam data, allow for the targeted removal of wisps, particularly from detectors where their presence is most pronounced.

To put these wisp maps into use, we created a mask to isolate regions with significant wisp signal, ensuring that only areas affected by wisps were corrected. This then minimizes the addition of noise in the images.  Subsequently, the identified wisps underwent a convolution process with a Gaussian kernel (with a standard deviation of 2 pixels) to smooth their appearance and reduce pixel noise. The smoothing was deliberately kept minimal, as the wisps exhibit significant small-scale structure; a 2-pixel kernel provided an optimal balance between enhancing the signal-to-noise ratio and preserving the fidelity of the wisp features for accurate subtraction from the original count-rate images.

The final phase of correction aimed to subtract the `wisps' from the images, minimizing the variance between the {\it image -- the wisp} template. This involved masking bright sources to prevent their interference and adjusting the scaling factor (see Fig.~3 in \citealt{Bagley2023}) of the convolved image to achieve the best match. Through careful calibration, we determined the optimal factor that, when applied, effectively removed the influence of wisps from the data, as depicted in the corrected images (Fig.~\ref{Fig::Wisps}).

\subsubsection{1/f Noise subtraction}
\label{subsec::1_f}

The subtraction of 1/$f$ noise, the correlated noise introduced during detector readout \citep[e.g.,][]{Schlawin2020}, was a crucial step in the NIRCam data processing, as the official JWST pipeline did not yet incorporate its removal at the time of processing. This effect manifests as faint horizontal and vertical striping across the images. Following an approach similar to that of \citet{Bagley2023}, we first masked astrophysical sources using a multi-tiered detection method that combines several Gaussian kernels to capture both extended and compact emission. On the masked images, we estimated and removed the residual background patterns by computing sigma-clipped medians along rows and columns. For horizontal striping, this was performed amplifier-by-amplifier to account for intra-detector variability. The original algorithm is described in detail in \citet{Bagley2023}.

\subsubsection{Correction of Persistence}
\label{subsec::persistence}
Persistence in the JWST detectors refers to the phenomenon whereby residual images of bright sources persist in subsequent exposures \citep{Rieke2023}. Persistence can significantly complicate the analysis of faint astronomical objects, as it may mimic astrophysical signals such as a Lyman break in galaxy spectra.

The underlying mechanisms of persistence have been the subject of various studies, with theoretical models proposed to explain its origins and behavior \citep{Smith2008}. Additionally, empirical analyses of NIRCam's flight detectors have further elucidated the persistence characteristics observed during operations \citep[e.g.,][]{Leisenring2016}

To avoid persistence artifacts that could mimic a Lyman break, for example, near a bright star when filters are observed from long to short wavelengths,  the sequencing of filter observations is carefully planned. Specifically, the F115W and F277W filters are employed prior to the F150W and F444W filters to minimize the likelihood of persistence affecting critical measurements; in other words, by observing in this sequence, persistence should fade pivoting from shorter wavelength exposures to longer wavelengths, and not the opposite, which could lead to certain persistence signals potentially masquarading as high-redshift Lyman-break galaxies. Nonetheless, to address instances where persistence may still occur, we have adopted the \texttt{PersistenceFlagStep} functionality from the \texttt{snowblind} package (\citealt{Davies2024}; v.0.2.1)\footnote{\url{https://github.com/mpi-astronomy/snowblind}}, a tool specifically designed to identify and mitigate persistence effects in JWST data.

The `snowblind' implementation introduces a dedicated step in the data processing pipeline, identifying any pixel that reaches saturation in exposure $N$ and applying a mask to these pixels in subsequent exposures ($N+n$), where $n$ is a time-dependent variable. This approach is based on the availability of the saturation mask found within the `GROUPDQ' extension of the data files, which does not automatically propagate to the `DQ' extension of the `*rate.fits' files. By leveraging this method, we ensure that pixels affected by persistence are more systematically flagged and excluded from further analysis, thereby preserving the integrity of the scientific results derived from the COSMOS-Web survey data.

\subsubsection{Correction of `Claw' Artifacts}
\label{subsec::claws}
\begin{figure}[!]
\centering
\includegraphics[width=0.4\textwidth]{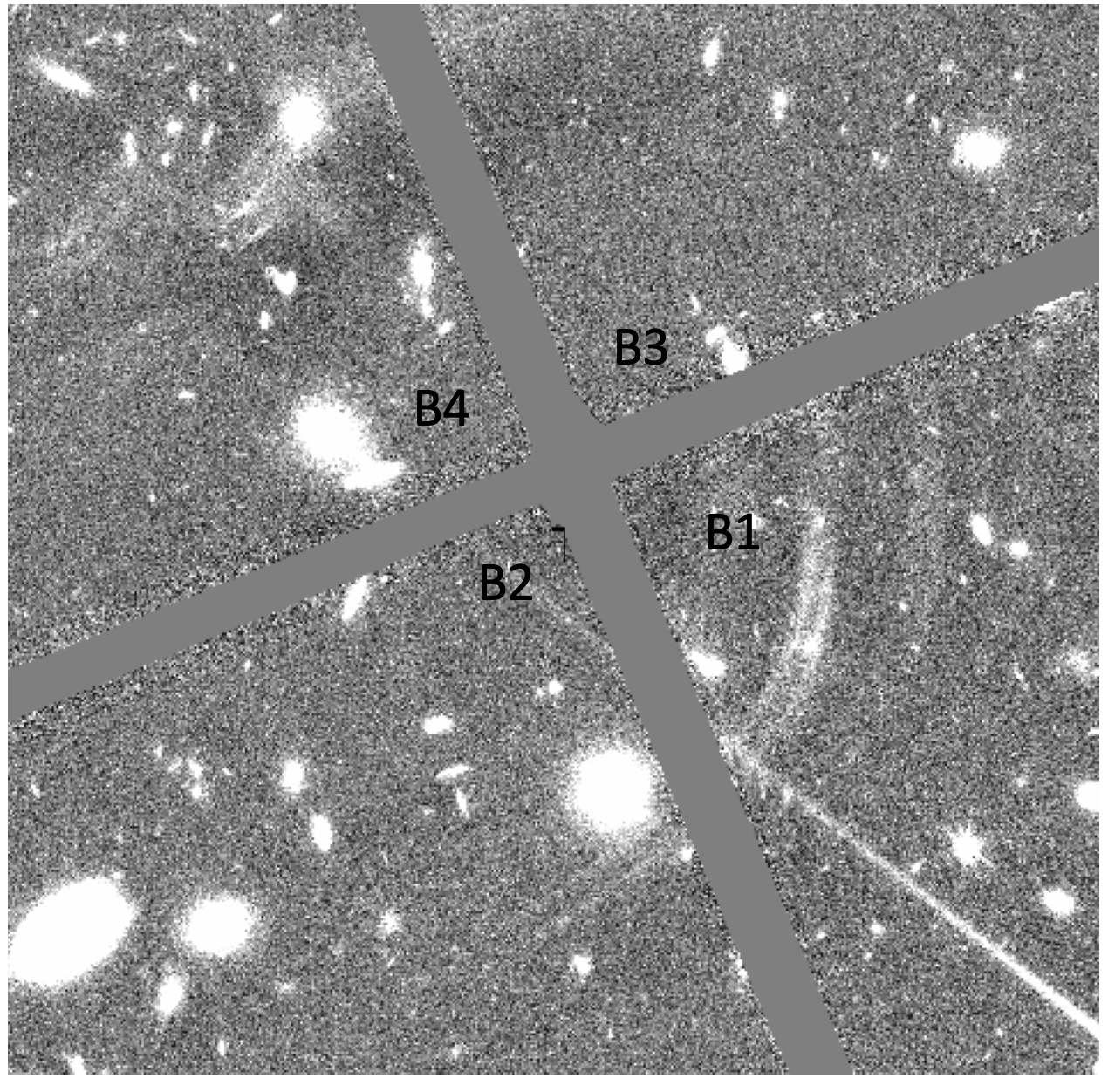} 
\caption{Zoom-in on module B of the F150W filter for a calibrated image before any correction. The 4 detectors are shown in the image, with a more pronounced claw shape for detector B1. The technique for subtracting these claws is presented in Section \ref{subsec::claws}
 and illustrated in Fig.~\ref{Fig::claws}.}         \label{Fig::claw_example}
\end{figure}

The presence of `claw' artifacts in NIRCam data, arising from the intricate interaction between the optical design and light's behavior on the detector array, poses significant challenges for data analysis, particularly when examining faint objects in proximity to brighter sources. These artifacts are predominantly observed in module B, especially within quadrant B1, with the F150W filter being most affected (Fig.~\ref{Fig::claw_example}). The variability of these artifacts in location and shape necessitated a customized approach for their effective mitigation to preserve data quality.

Addressing the `claws' required a detailed method that combines visual inspection with specialized computational techniques due to their inconsistent intensity and appearance. The first step in this approach involved the careful visual identification and delineation of the artifacts.

\begin{figure}[!]
\centering
\includegraphics[width=0.48\textwidth]{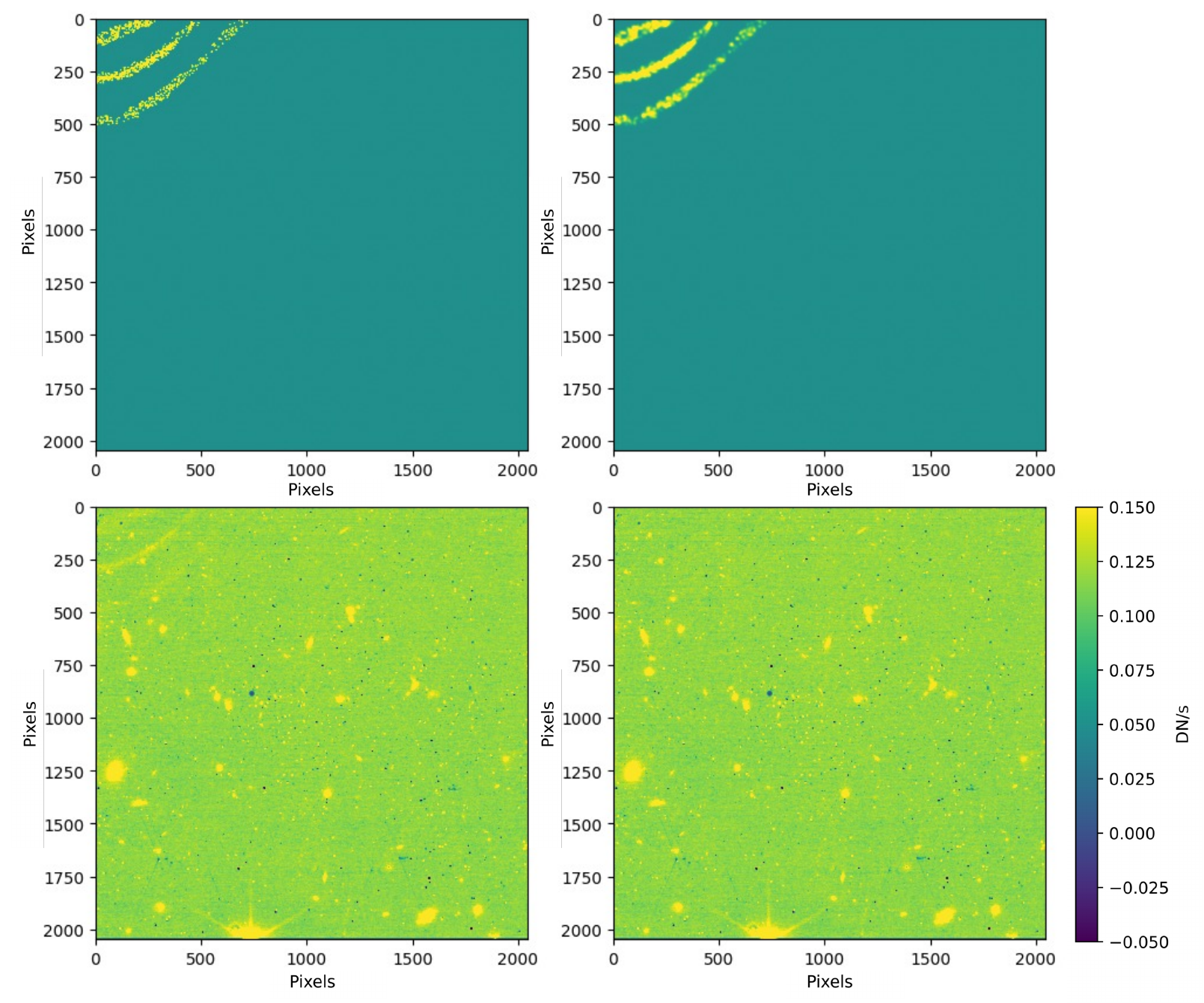} 
\caption{Different steps for claw subtraction. The claws were isolated from the image (top-left), then smoothed (top-right), then subtracted from the original image (bottom-left), minimizing variance to obtain the final image (bottom-right).}
         \label{Fig::claws}
\end{figure}

This step was implemented using the following publicly available script (Crab.Toolkit.JWST)\footnote{\url{https://github.com/1054/Crab.Toolkit.JWST}}, which is specifically designed to facilitate this process by isolating and removing ‘claws’ artifacts from the count-rate images (*rate.fits). After identifying the regions affected by ‘claws’, the script generates an image isolating these artifacts (top left panel in Fig.~\ref{Fig::claws}). This image is subsequently convolved with a two-dimensional Gaussian kernel with a standard deviation of 6 pixels (optimal trade-off between enhancing the signal-to-noise ratio and preserving the fine-scale structure of the claws signal), effectively smoothing the ‘claws’ signal and facilitating its treatment in subsequent processing steps (top right panel in Fig.~\ref{Fig::claws}).

The subsequent step involves minimizing the variance between the original count-rate image, where bright sources have been masked to avoid interference, and the convolved claws image. The masking process involves setting a detection threshold to identify and exclude bright sources effectively. By adjusting the scaling factor for the convolved image between 0 and 2, we aim to reduce the variance to a minimum. Determining the optimal scaling factor allows for the subtraction of the scaled convolved claws image from the original image, resulting in a corrected image less affected from these artifacts (bottom right panel in Fig.~\ref{Fig::claws}).

The January 2023 (Jan23) data exhibited pronounced ‘claw’ artifacts, which we actively removed from the affected images. The April 2023 (Apr23) observations were more favorable in this regard, as no significant ‘claws’ were present, and no correction was required. In the January 2024 (Jan24) dataset, while the artifacts were still detectable, they were sufficiently faint that we opted not to remove them to avoid the risk of suppressing faint astrophysical signals in the affected regions.

\subsection{Pipeline Level 2}

Following the initial corrections applied in Stage 1, we further processed the count-rate images using the Stage 2 pipeline, specifically the \texttt{Image2Pipeline}, which was executed mainly with the default settings (modifications to the default configuration are described in the following section). This stage involves the assignment of World Coordinate System (WCS) information, flat-fielding, and photometric calibration, culminating in fully calibrated individual exposures. Each rate map from Stage 1 is thus transformed into a calibrated science image, with units converted from counts per second to MJy/sr. We added only one step to this stage, the subtraction of a global sky background offset (Section~\ref{subsec::global_sky_background}).

\subsubsection{Subtraction of a global offset}
\label{subsec::global_sky_background}
An additional step implemented at this stage is the subtraction of a global sky background offset present in the images. This procedure, inspired by the technique employed by \cite{Bagley2023}, involves masking bright sources within the images before subtracting the offset. This offset is precisely determined by fitting a Gaussian to the overall distribution of pixel values for each detector, thereby ensuring the removal of any global background signal and refining the quality of the calibrated science images.

\begin{figure*}[!]
\centering
\includegraphics[width=1\textwidth]{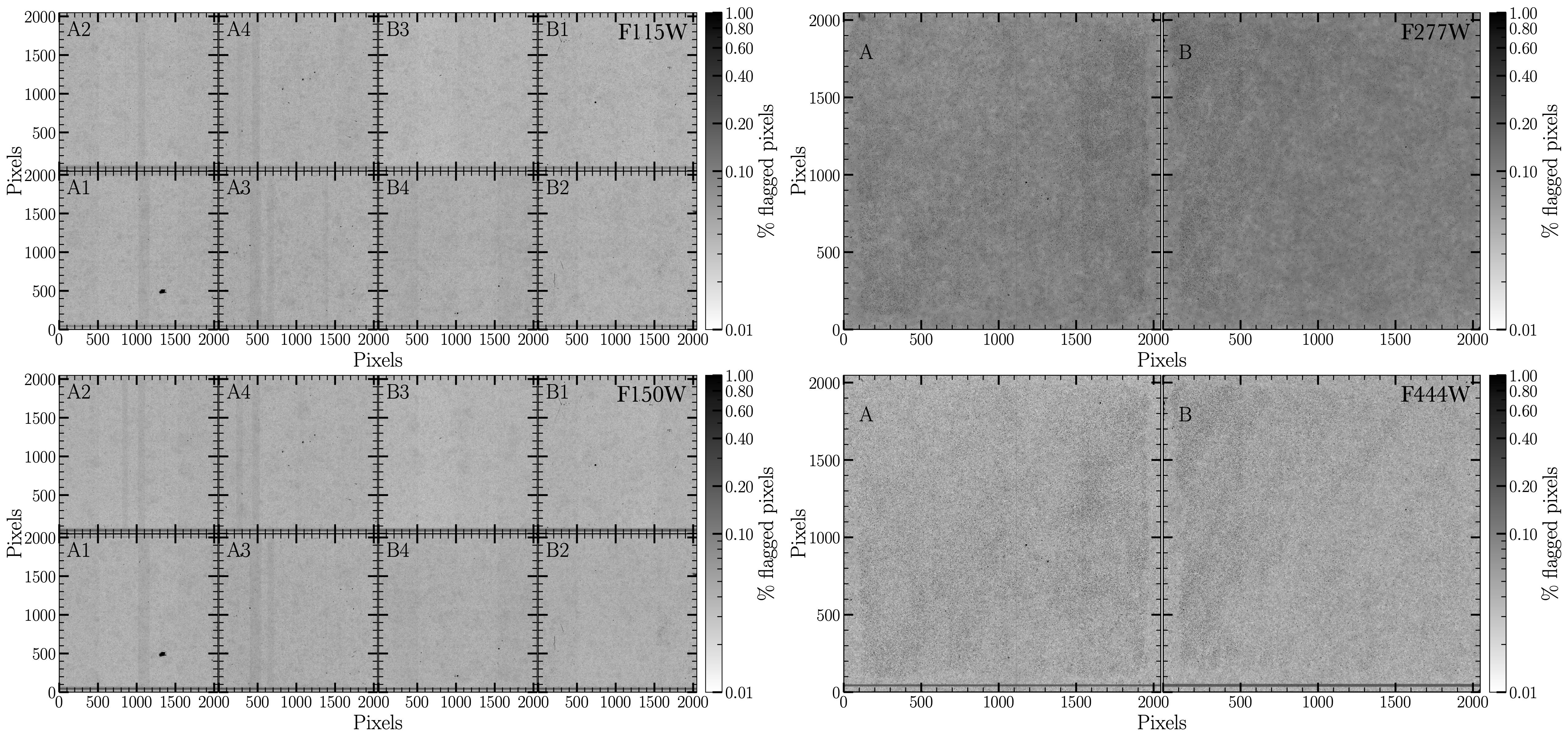} 
\caption{For each module and detector, this figure illustrates the percentage of pixels exhibiting a non-zero value in the Data Quality (DQ) array across all calibrated files from April 2023. Pixels were flagged in the following of the data reduction process when the proportion of non-zero DQ array values exceeded 20\% (pixels that are bad in >20\% of the exposures), indicating potential issues with data quality. The percentage of such flagged pixels is detailed in Table~\ref{tab::flagged_pixels}.}
         \label{Fig::bad_pixels}
\end{figure*}

\subsection{Pipeline Level 3}

In the third step of the JWST pipeline, we combined the different exposures into a unified mosaic. This phase encompasses astrometric alignment, background normalization, outlier detection, and a resampling step. This section details the methodologies and parameters employed in this step to create the final mosaic. Each of these steps has been executed independently to facilitate computing memory optimization, thereby conserving essential memory resources for the subsequent data processing phases.

\begin{deluxetable}{ccccc}
\tabletypesize{\scriptsize}
\tablecaption{Percentage of Pixels Flagged More Than 20\% of the Time by Detector and Filter\label{tab::flagged_pixels}}
\tablewidth{0pt}
\tablehead{
\colhead{Filter}  & \colhead{Detector} & \colhead{\% Flagged} & \colhead{Detector} & \colhead{\% Flagged}
}
\startdata
F115W & A1 & 2.12\% & B1 & 1.88\% \\
F115W & A2 & 1.77\% & B2 & 1.91\% \\
F115W & A3 & 1.90\% & B3 & 1.69\% \\
F115W & A4 & 1.80\% & B4 & 1.84\% \\
F150W & A1 & 2.97\% & B1 & 2.77\% \\
F150W & A2 & 2.21\% & B2 & 2.48\% \\
F150W & A3 & 2.70\% & B3 & 2.25\% \\
F150W & A4 & 2.43\% & B4 & 2.55\% \\
F277W & A  & 5.89\% & B  & 4.66\% \\
F444W & A  & 5.85\% & B  & 4.63\% \\
\enddata
\tablecomments{This table summarizes the percentage of pixels flagged as defective more than 20\% of the time for each detector and filter combination. The assessment was based on the analysis of the data quality (DQ) extension of calibrated (*cal.fits) files from April 2023. The flagged percentage indicates the proportion of pixels identified as potentially defective across different tiles and detectors.}
\end{deluxetable}

\subsubsection{Identification and Management of Bad Pixels}

The presence of ``hot" or bad pixels poses significant challenges, particularly in the study of galaxies during the epoch of reionization, where sources are typically compact and detected in only a limited number of filters. In such cases, the loss or corruption of even a small number of pixels can substantially impact photometric measurements and source characterization. Given the relatively limited overlap between different visits in the COSMOS-Web survey, a critical pre-processing step was incorporated prior to executing the \texttt{OutlierDetection} stage of the JWST pipeline's Stage 3. This step involved a statistical evaluation to identify and exclude pixels with the highest probability of being faulty.

To accomplish this, we stacked all calibrated files (*cal.fits) from the Apr23 dataset. Each detector and filter was stacked separately to ensure a detailed assessment. Within each calibrated file, we examined the data quality (DQ) extension to identify pixels that were flagged more than 20\% of the time across all observations, irrespective of the specific flag value.  This 20\% threshold was empirically determined to optimize the identification of truly defective pixels: lower thresholds began to pick up spurious flags due to noise or transient effects, while higher thresholds missed a significant fraction of consistently problematic pixels. Such pixels were considered defective and were accordingly flagged in the DQ array of the corresponding calibrated images.

The percentage of pixels flagged as defective, differentiated by detector and filter, is illustrated in Fig.~\ref{Fig::bad_pixels}. This represents between 1.7\% and 3.0\% of the pixels per sensor, for the short wavelength detectors, that are flagged, with no major difference between the two modules (2.2\% of flagged pixels on average for modules A and B). For the long wavelength detectors, the percentage of flagged pixels is a little higher, with between 4.6 and 5.9\% of the pixels being flagged. The positions of the flagged pixels are made available as part of our data release.

\subsubsection{Astrometric Calibration of NIRCam tiles}
\label{subsec::astrometry}
Achieving accurate absolute and relative astrometry in the JWST mosaics across all filters is crucial for the integrity of subsequent measurements, encompassing photometry, morphology, the estimation of photometric redshifts, or the search for transient phenomena. To this end, we employed the external JWST/HST Alignment Tool \citep[JHAT;][]{Rest2023} procedure\footnote{\url{https://jhat.readthedocs.io/en/latest/}} version 0.0.1 for astrometric calibration, finding it to offer enhanced precision over the JWST's native TweakReg procedure, which is integrated within the JWST data processing pipeline.

For the astrometric alignment process, we constructed a reference catalog for the COSMOS-Web region, using a newly created 0.03"/pixel mosaic derived from the original COSMOS HST/F814W imaging data \citet{Koekemoer2007}. This new version of the F814W mosaic had been reprocessed to align with the latest astrometric standards, following the approaches originally described in \citet{Koekemoer2011}, including a direct astrometric alignment to Gaia-DR3 \citep{gaia21} as well as the COSMOS2020 catalog \citep{Weaver2022}, thereby ensuring superior astrometric accuracy. From the reference catalog created from these new F814W mosaics, we excluded stars that could potentially bias the astrometry due to their proper motion. This was achieved by removing objects with a stellarity index measured by \texttt{SExtractor} \citep{Bertin1996} greater than 0.85, a size smaller than 5 pixels, or a magnitude brighter than 19 AB, thereby refining the catalog for optimal alignment with our NIRCam data. 

\begin{figure*}
\centering
\includegraphics[width=0.95\textwidth]{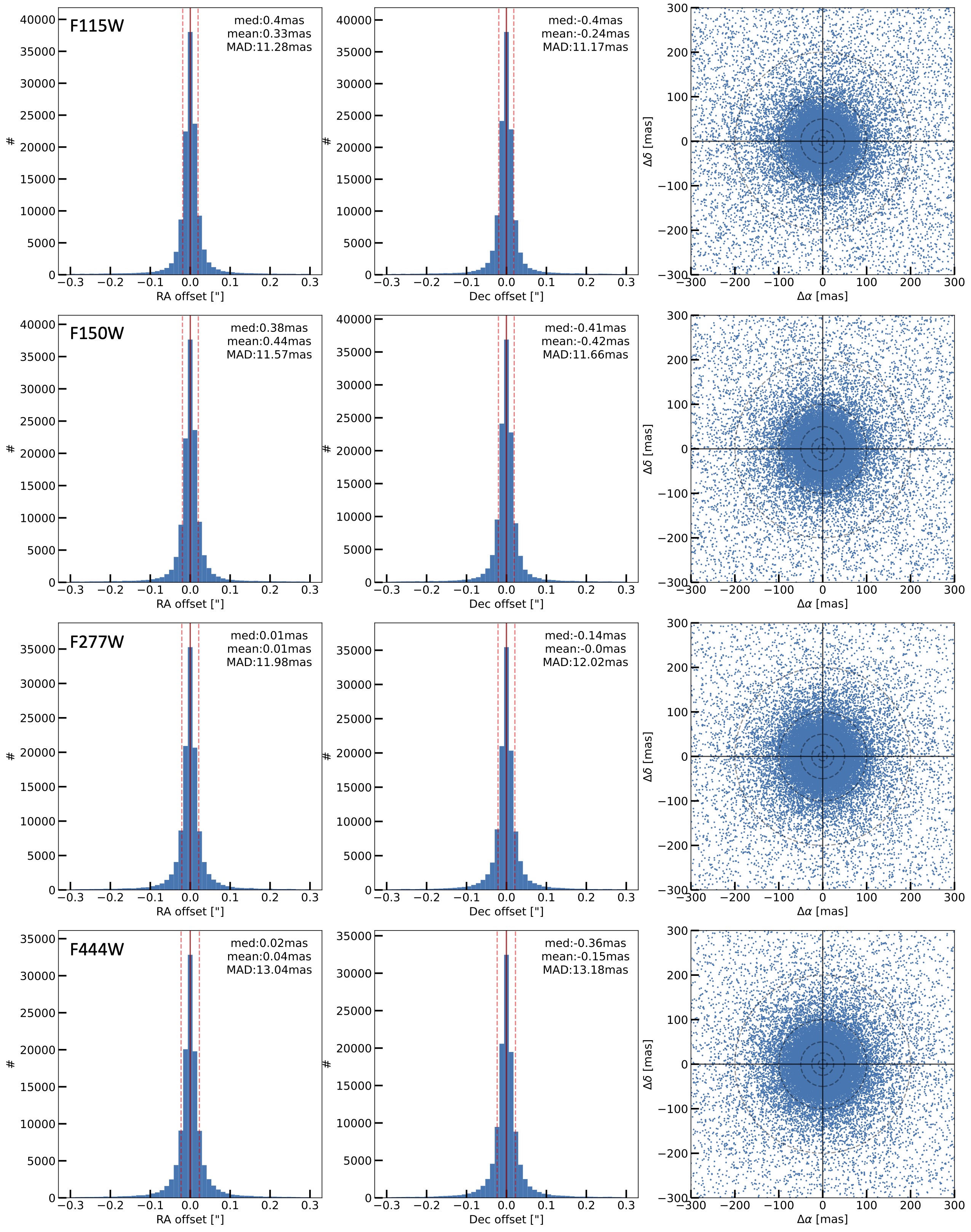} 
\caption{Diagnostic plots illustrating the astrometric alignment in COSMOS-Web. Results for other individual tiles are provided in the Appendix, while aggregate results for all tiles are detailed in Table~\ref{tab::astrometry_total}. Across all the tiles, neither the average nor median offset exceeds 3 mas in RA or 4 mas in Dec, with the Median Absolute Deviation (MAD) remaining below 14 mas.}
         \label{Fig::astrometry_histograms}
\end{figure*}

Moreover, the reference catalog was specifically constructed to align the images observed with the LW filters, as these detectors cover a larger area and the corresponding images are deeper. This results in a higher number of reliable matches to the HST reference catalog. Following this, the F277W tiles were employed as a basis to align the SW filters, ensuring consistency across the entire spectral range of NIRCam observations. This alignment process was facilitated by an automated version\footnote{\url{https://github.com/1054/Crab.Toolkit.SExtractorPlus}} of \texttt{SExtractor}. For each mosaic, we chose the windowed centroid coordinates of the detections, \texttt{XWIN\_IMAGE} and \texttt{YWIN\_IMAGE}, which we transformed into celestial coordinates using the World Coordinate System (WCS) embedded within the mosaics. This approach successfully accounts for the distortions present in the original images. Across all tiles, both the average and median offsets relative to the reference catalog remain below 3 mas in RA and 4 mas in Dec, with the Median Absolute Deviation (MAD) staying below 14 mas, successfully accounting for the distortions present in the original images (Fig.~\ref{Fig::astrometry_histograms}).

\subsubsection{Background removal}
\label{subsec::background}

Accurate background subtraction is essential, particularly for the detection of faint and distant galaxies. Insufficient removal of background signals can significantly impair data integrity, influencing photometric precision, faint object identification, and the reliability of subsequent analyses. We perform the background removal in the *crf.fits files just before the resampling step.

In our approach, bad pixels are initially identified and masked based on data quality (DQ) indicators to ensure they do not distort the background estimation. We then proceed to compute an initial background estimate by implementing a two-dimensional polynomial fit after masking regions representing the upper 70th percentile of brightness and applying a median filtering technique, thereby mitigating the influence of bright sources.

The procedure continues with iterative refinements: employing sigma clipping techniques to progressively mask fainter sources, each iteration expanding the masks using a circular top-hat kernel with a decreasing radius, tailored to capture to increasingly subtle intensity variations within the image.

Following the construction of these masks, we conduct a background subtraction using the \texttt{Astropy/Photutils} \texttt{Background2D} class. This involves adopting the \texttt{MedianBackground} estimator within sigma-clipped regions to accurately determine the background across 17x17 pixel boxes, applying a 3x3 pixel filter to smooth the estimated background levels.

\begin{figure*}
\centering
\includegraphics[width=1.\textwidth]{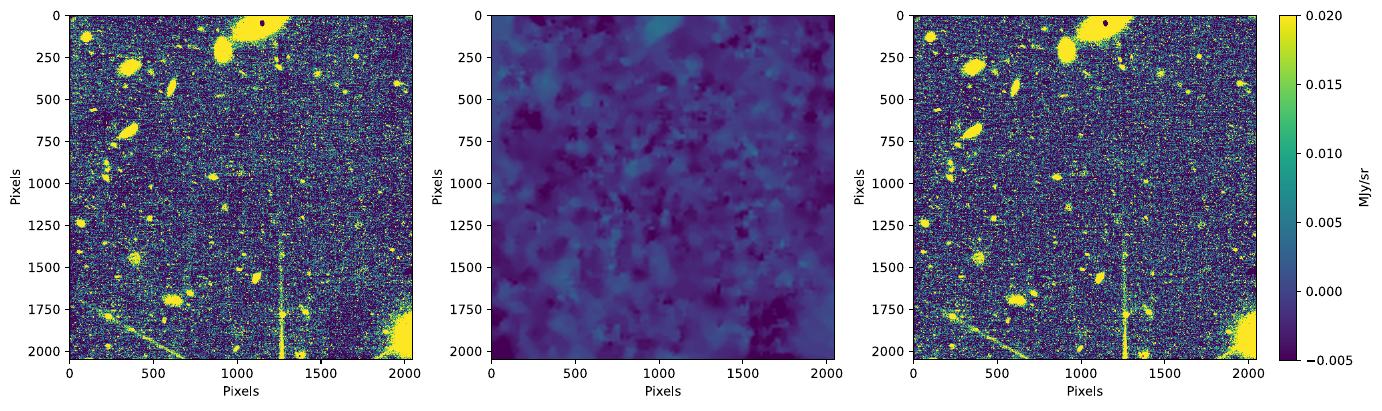} 
\caption{Illustration of the background subtraction process applied to the final pre-mosaic images (*crf.fits files). The left image displays one of the final images before mosaic assembly, serving as the initial state for background subtraction. The middle image depicts the extracted background, as detailed in Section~\ref{subsec::background}, highlighting the method used to isolate the global background signal. The right image shows the result after the subtraction of this background from the initial image.}
         \label{Fig::background}
\end{figure*}

Additionally, a pedestal level adjustment is made to correct for any constant offsets, utilizing a Gaussian fit to the background-subtracted, masked data. An illustrative example displaying the original image, extracted background, and the background-subtracted image is provided in Fig.~\ref{Fig::background}.

\subsubsection{Outlier Rejection}

We then perform a \texttt{sky\_match} step. This is conducted using the default parameters of the JWST pipeline to ensure uniform sky levels across all exposures. Subsequent to this sky matching, we proceed with the Outlier Rejection step in preparation for the final mosaic assembly.

While the majority of cosmic rays are flagged in Stage 1 of the JWST pipeline, additional measures are necessary to address the cosmic rays that evade initial detection and other defective pixels (such as dead, hot, or noisy pixels). To mitigate this, we run the \texttt{OutlierDetectionStep} of JWST pipeline Stage 3, focusing on further refining the data quality.

For each filter and visit within the COSMOS-Web survey, we compiled all the images (*jhat.fits) overlapping this visit into an association file (ASN), which serves as input for the Outlier Detection step. Despite the relatively sparse overlap between different visits in the COSMOS-Web context, this approach ensures thorough outlier flagging in regions of intersection.

To optimally identify outliers in areas with limited exposure counts, we adopted the following parameters:
\begin{itemize}\setlength\itemsep{-0.5em}
\item \texttt{maskpt = 1}
\item \texttt{nlow = 0}
\item \texttt{nhigh = 1}
\item \texttt{pixfrac = 1}
\item \texttt{kernel = ‘square'}
\end{itemize}

After conducting multiple trials, we adopted a \texttt{pixfrac = 1} setting. Given that the sub-pixel phase space in our mosaics is typically sampled by only a few exposures, smaller \texttt{pixfrac} values would result in substantial inhomogeneities in the resampled images. Choosing \texttt{pixfrac = 1} ensures more uniform sub-pixel coverage across the field, while still balancing sensitivity to outliers and the preservation of genuine astronomical signals.

\subsection{Resampling Step}

Due to the extensive memory requirements for processing the entire COSMOS-Web data set simultaneously, we did the resampling step individually in each of the 20 tiles (see Section~\ref{subsec::tile}), to facilitate more manageable processing loads.

In constructing each tile, we employed the Stage 3 Resample routine, which combines all dithered exposures into a final mosaic. This process involves selecting all overlapping images (*crf.fits) for a given tile and adding them to an association file (ASN), which then serves as the input for the resampling step.

For each tile, the tangent point remains constant, set to \texttt{CRVAL1} = 150.1163213 and \texttt{CRVAL2} = 2.200973097, aligning with the central coordinates of the COSMOS field as established in previous studies \citep{Koekemoer2007, Koekemoer2011, Laigle2016, Weaver2022}.

Each tile is produced in three resolutions, 20 mas, 30 mas and 60 mas, to accommodate different scientific needs and computational constraints. The 20 mas mosaics are primarily used for weak lensing analysis (Scognamiglio et al. in preparation), while the 30 mas and 60 mas mosaics are more widely used in the team for bulk photometric and morphological galaxy measurements. The geometric configuration of each tile adheres to a rectangular shape with a 20-degree inclination, measuring 9600 by 12455 pixels at the 60 mas resolution, doubling (tripling) in each dimension for the 30 mas (20 mas) resolution.

\section{Results}
\label{sec:results}

\subsection{Data Release}
\label{subsec::data_release}
The reduced data from the COSMOS-Web survey are publicly accessible via the following URL: \url{https://exchg.calet.org/cosmosweb-public/DR1/}. In our commitment to scientific transparency and advancement, we regularly update the dataset to incorporate the latest improvements and refinements in our data reduction processes. A detailed README file accompanies these updates, providing comprehensive information on the data reduction advancements and modifications.

The data for the 20 tiles (ranging from A1 to A10 and B1 to B10) are provided in the ``i2d.fits" format. These ``.i2d.fits" files contain seven extensions beyond the primary header: \texttt{SCI}, \texttt{ERR}, \texttt{CON}, \texttt{WHT}, \texttt{VAR\_POISSON}, \texttt{VAR\_RNOISE}, and \texttt{VAR\_FLAT}, with each extension's definition detailed in the JWST documentation\footnote{\url{https://jwst-pipeline.readthedocs.io/en/latest/jwst/data_products/science_products.html}}. To facilitate data usability, the \texttt{SCI}, \texttt{ERR}, and \texttt{WHT} extensions have been extracted and organized within a dedicated sub-folder named `extension\_mosaics`. Each mosaic is compressed and available in two resolutions, featuring pixel scales of 30 mas and 60 mas, to aid in data manipulation and facilitate download efficiency. The 20 mas mosaics are available upon request.

\begin{figure*}[htbp]
\centering
\includegraphics[width=1\textwidth]{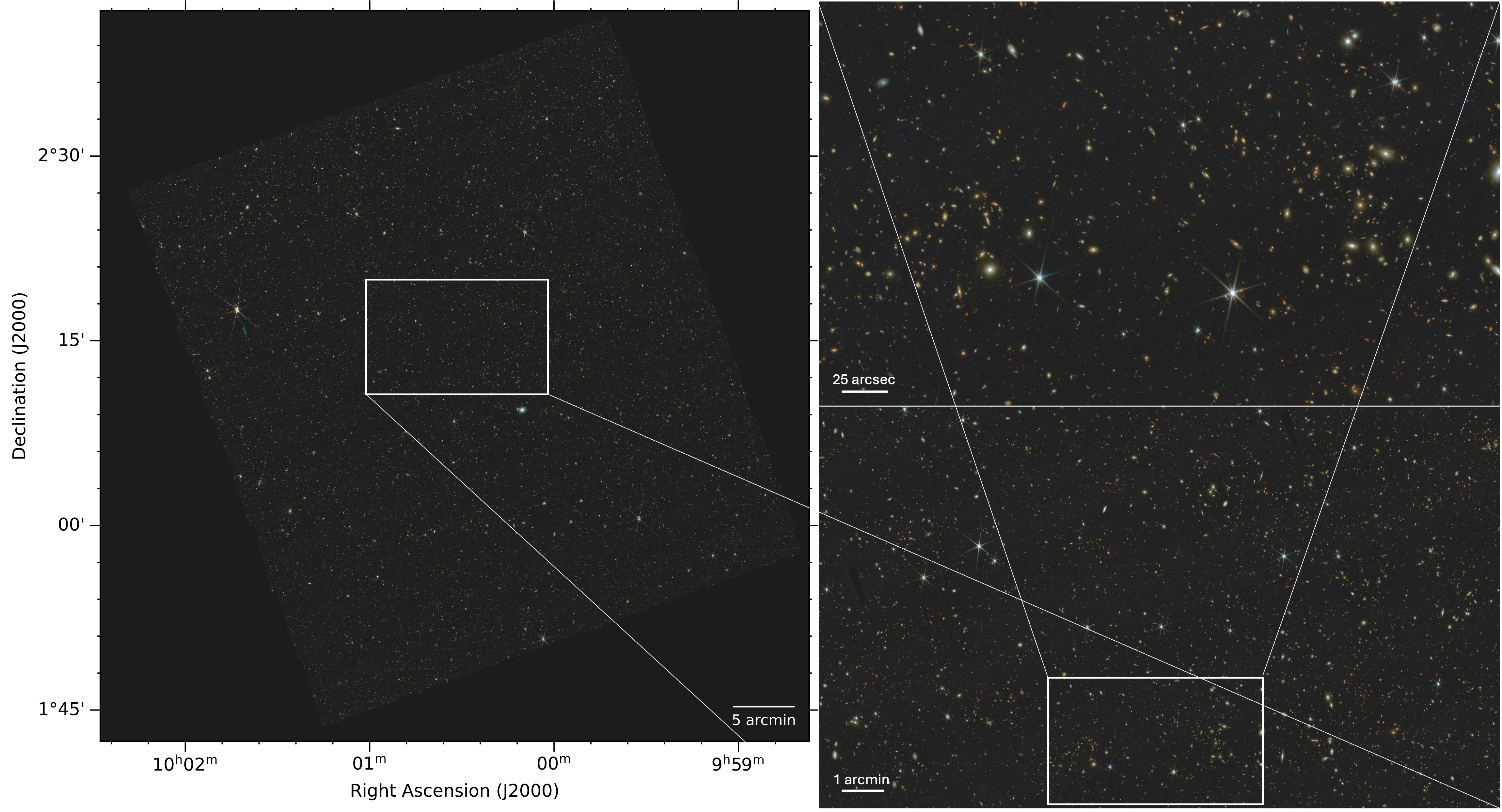} 
\caption{Four-color (F115W, F150W, F277W, F444W) image of the NIRCam COSMOS-Web mosaic with 2 progressive "zoom-in"s inside a random region in this field. A full resolution version of this image can be found in the same repository as the data release. The color image was made using the \texttt{Trilogy} software package \citep{Coe2012}.}

         \label{Fig::RGB_image}
\end{figure*}

In addition to these FITS images, we provide a color image synthesized from the four NIRCam filters (see Fig.~\ref{Fig::RGB_image}) using the \texttt{Trilogy} software package\footnote{\url{https://github.com/dancoe/trilogy}} \citep{Coe2012}. A mask is also supplied in both 30 mas and 60 mas resolutions, indicating the number of NIRCam exposures associated with each pixel.

\subsection{Survey Depth}

While \citet{Casey2023} presented depth estimates for the first six pointings using pipeline version 1.8.3 and CRDS pmap 1017, our full-survey analysis used updated in-flight calibrations. The depth remains uniform across the survey footprint, with a standard deviation below 0.04 AB magnitudes when comparing regions with identical exposure counts. Table~\ref{tab::depth} summarizes the achieved depths in each of the four NIRCam filters as a function of the number of exposures. Depths were estimated by measuring the flux within 100,000 randomly placed circular apertures of 0.15" radius for each exposure tier. Apertures crossing boundaries between regions of differing coverage were excluded. The flux distribution was fitted with a Gaussian, and the full width at half maximum (FWHM) was multiplied by the desired significance level (here, 5$\sigma$), without applying aperture corrections.

These findings are in agreement with JWST's anticipated in-flight performance metrics \citep{Rigby2023}, and in fact show slightly better sensitivity than expected. They also correlate well with results from detailed simulations of the COSMOS-Web field based on the DREaM semi-empirical model \citep{Drakos2022}, which will be further discussed in Drakos et al. (in preparation).

\begin{deluxetable}{ccccc}
 \tabletypesize{\scriptsize}
 \tablecaption{COSMOS-Web NIRCam Survey Depths  \label{tab::depth}}
 \tablehead{
 \colhead{Nb. of} & \colhead{F115W}  & \colhead{F150W}  & \colhead{F277W}  & \colhead{F444W}\\[-0.25cm]
Exposures & 5$\sigma$ depth & 5$\sigma$ depth&5$\sigma$ depth& 5$\sigma$ depth
 }
\startdata
1 & 26.69 & 26.95 & 27.74 & 27.60\\
2 & 27.05 & 27.30 & 28.01 & 27.88\\
3 & 27.24 & 27.49 & 28.19 & 28.07\\
4 & 27.41 & 27.65 & 28.34 & 28.22\\
\enddata
\tablecomments{This table presents the 5$\sigma$ depths achieved across the COSMOS-Web NIRCam survey, segmented by the number of exposures. Depths were determined through a Gaussian fit to the flux measurements within 0.15" radius apertures.}
\end{deluxetable}

\begin{deluxetable*}{cccccccc}
\tabletypesize{\scriptsize}
\tablecaption{Absolute astrometry\label{tab::astrometry}}
\tablewidth{0pt}
\tablehead{
\colhead{Tile}  & \colhead{Filter} & \colhead{median $\Delta$RA} &  \colhead{mean $\Delta$RA} &  \colhead{MAD RA} & \colhead{median $\Delta$Dec} &  \colhead{mean $\Delta$Dec} &  \colhead{MAD Dec} \\[-0.25cm]
 & & (mas) & (mas) & (mas) & (mas) & (mas) & (mas) 
}
\startdata
All & F115W &  0.40   &   0.33  &   11.28 &   -0.40  &  -0.24 & 11.17  \\
All & F150W &  0.38   &   0.44  &   11.57 &   -0.41  &   -0.42 & 11.66  \\
All & F277W &  0.01   &   0.01  &   11.98 &   -0.14  &   0.00 & 12.02  \\
All & F444W &  0.02   &   0.04  &   13.04 &   -0.36  &   -0.15 & 13.18  \\
\enddata
\tablecomments{Summary of astrometric offsets observed between NIRCam detections across all four filters and the reference catalog for the entirety of the 20 tiles within the COSMOS-Web survey, as detailed in Section~\ref{subsec::astrometry}. The LW observations were directly aligned using the reference catalog, whereas an intermediary catalog based on the F277W filter observations was employed to align the SW exposures. Detailed results for each tile are provided in the Appendix in Table~\ref{tab::astrometry_total}.}
\end{deluxetable*}

\subsection{Astrometric Accuracy and Distortion}
The alignment of our NIRCam observations with the reference catalog demonstrated median positional offsets in both RA and Dec of less than 5 mas, regardless of the filter used. Furthermore, the median absolute deviation (MAD) across the survey field remained below 14 mas, with a low variation observed between different filters, indicating a high level of astrometric precision across the COSMOS-Web survey. Table~\ref{tab::astrometry} presents the total astrometric discrepancies between objects detected in the COSMOS-Web survey and those in the reference catalog. In the Appendix, we present these astrometric discrepancies for all the filters and all the tiles in Table~\ref{tab::astrometry_total}.

\begin{figure}[!]
\centering
\includegraphics[width=.49\textwidth]{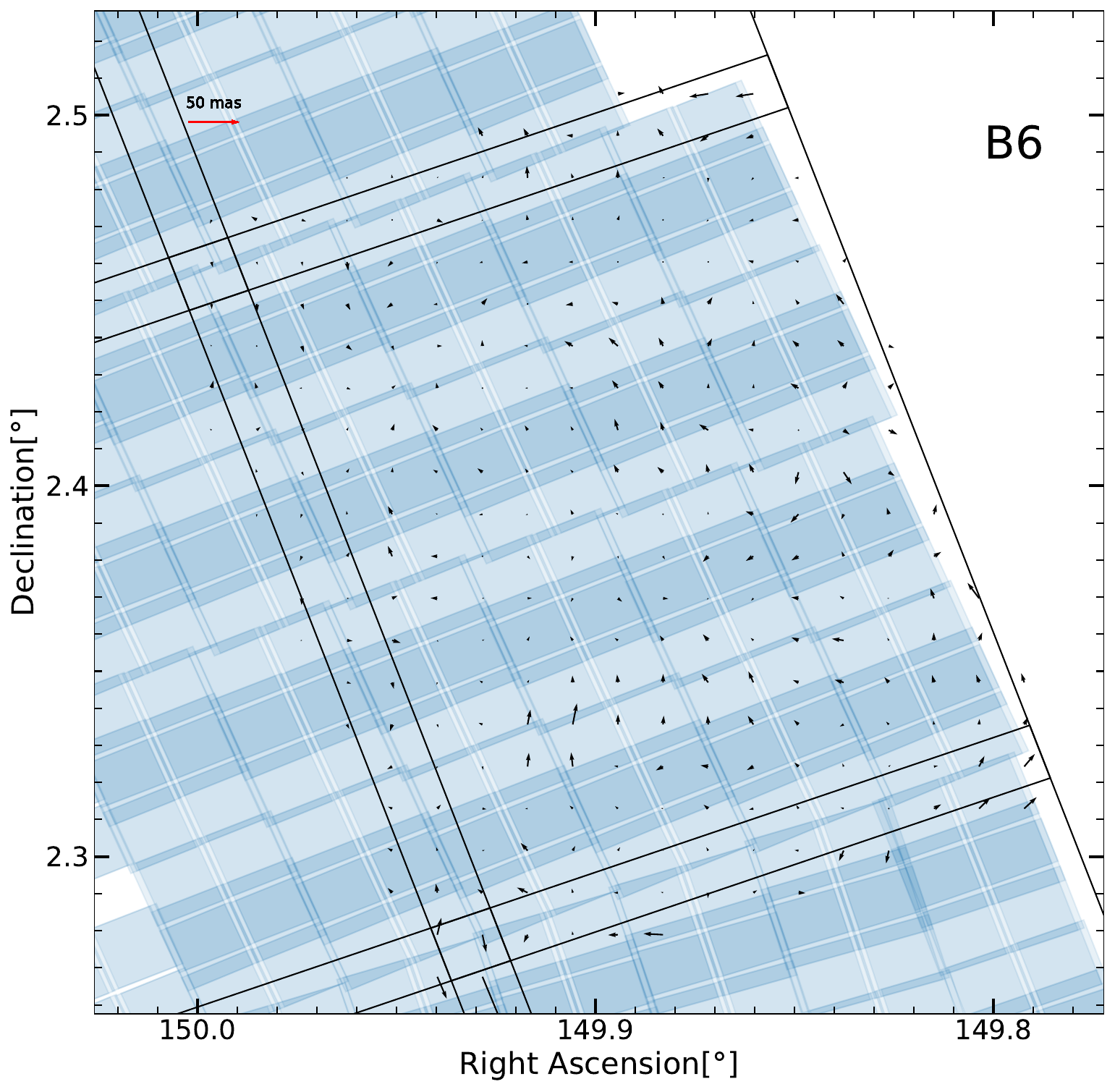} 
\caption{Example of the distortion map for the tile B6. Each arrow represents a sliding median including on average 35 galaxies, with an overlap of 40 percent between neighboring arrows. }
         \label{Fig::astrometry_vectors}
\end{figure}

Moreover, we conducted a comprehensive analysis to ensure consistency across the different detectors for each filter within the JWST NIRCam instrument. This verification process involved comparing the astrometric results obtained from each detector, thereby ensuring uniformity in the data quality and reliability of the astrometric calibration across the entire field of view. Additionally, we verified the absence of spatially coherent patterns larger than a few mas in the astrometric differences between the JWST observations and the reference catalog, shown in Fig.~\ref{Fig::astrometry_vectors}. Each arrow in the figure represents the median offset for a sliding median containing, on average, 35 sources, confirming the uniformity of our astrometric calibration across the survey area.

Relative astrometry between different filters, particularly among the LW and SW filters, revealed minimal discrepancies, underscoring the robustness of our astrometric calibration across the NIRCam instrument suite. The offset in astrometry between all the filters is less than 1 mas, with a MAD of $\sim$ 8 mas between LW filters, $\sim$ 9 mas between SW filters and up to $\sim$12 mas between LW and SW filters. The inter-filter astrometric differences are detailed in Table~\ref{tab::astrometry_relative}.

\begin{deluxetable*}{ccccccc}
\tabletypesize{\scriptsize}
\tablecaption{Relative Astrometry between the different filters of COSMOS-Web\label{tab::astrometry_relative}}
\tablewidth{0pt}
\tablehead{
 \colhead{Filter} & \colhead{median $\Delta$RA} &  \colhead{mean $\Delta$RA} &  \colhead{MAD RA} & \colhead{median $\Delta$Dec} &  \colhead{mean $\Delta$Dec} &  \colhead{MAD Dec} \\[-0.25cm]
 & (mas) & (mas) & (mas) & (mas) & (mas) & (mas) 
}
\startdata
F444W - F277W &  -0.01   &   0.01  &   7.45 &   0.22  &   0.32 & 7.30  \\
F444W - F150W &   0.27   &   0.39  &  10.33 &   0.20  &   0.37 & 10.24  \\
F444W - F115W &   0.27   &   0.33  &  11.46 &   0.17  &   0.21 & 11.29  \\
F277W - F150W &   0.38   &   0.47  &   8.98 &  -0.09  &  -0.09 & 8.77   \\
F277W - F115W &   0.27   &   0.35  &  10.02 &  -0.11  &  -0.17 & 9.91  \\
F150W - F150W &  -0.07   &  -0.23  &   8.38 &   0.04  &  -0.15 & 7.96   \\
\enddata
\tablecomments{For each combination of two filters in COSMOS-Web, we analyzed their relative astrometric difference by calculating the median, mean and the Median Absolute Deviation (MAD) in RA and Dec. Regardless of the combination of two filters examined, the discrepancy in either RA or Dec is less than 1 mas, and  MAD is under 12 mas. This MAD is minimized when comparing the Long Wavelength (LW) filters with each other, followed by the comparisons among Short Wavelength (SW) filters.}
\end{deluxetable*}

\subsection{Remaining Artifacts}
\label{sec:remaining_artifacts}

\begin{figure*}
\centering
\includegraphics[width=.8\textwidth]{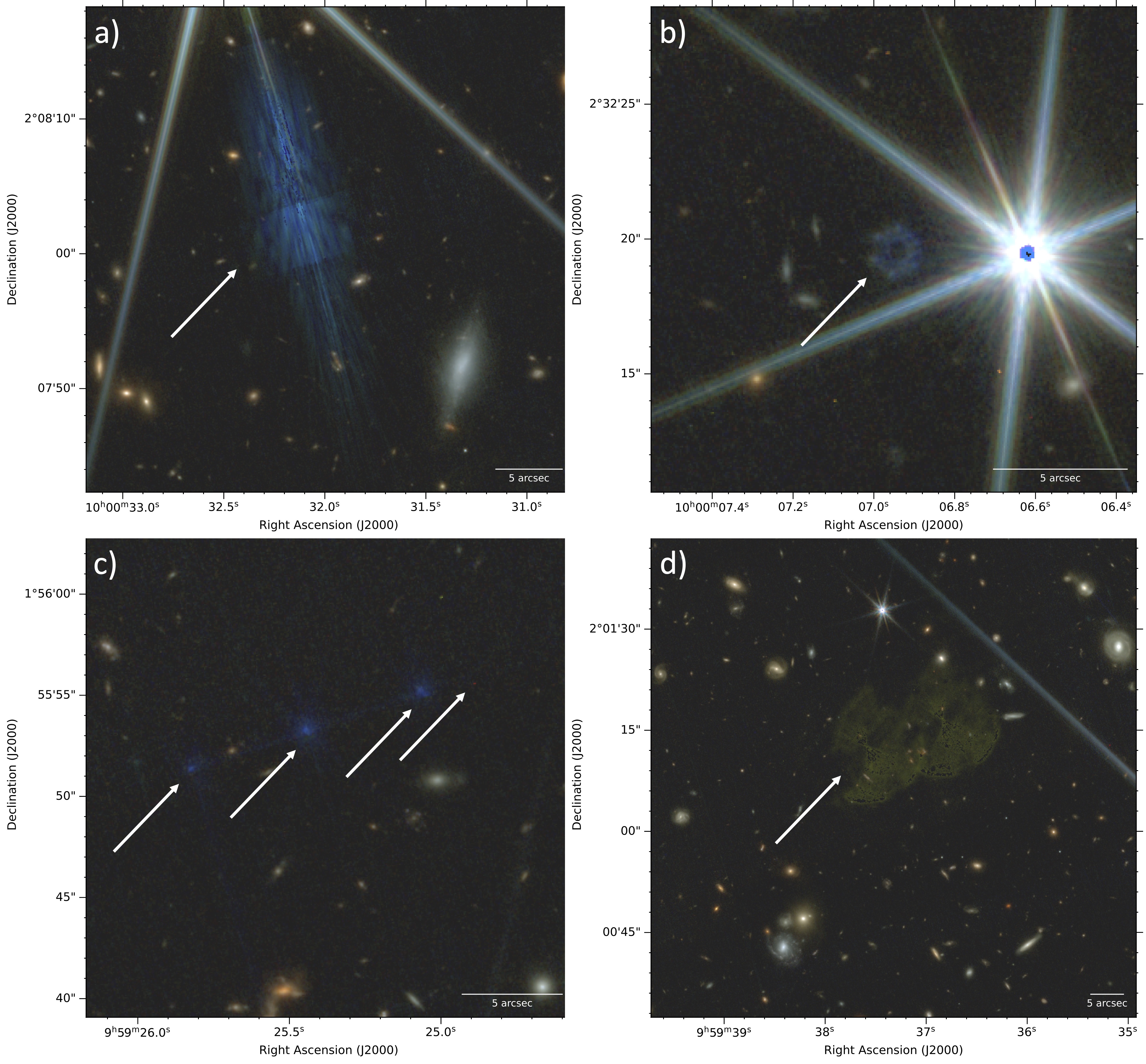} 
\caption{Examples of various artifacts remaining in the COSMOS-Web imaging. a) Dragon's Breath Type II, b) ghost star, c) persistence and hot pixel (right arrow), d) scattered light artifacts close to a bright star. These features, described in Section~\ref{sec:remaining_artifacts}, are currently under investigation and will be addressed in future COSMOS-Web data releases.}
         \label{Fig::artifacts}
\end{figure*}

This section addresses further features and anomalies identified in the NIRCam data, which have been deferred for treatment in a subsequent data release. Among these are the ‘Dragon's Breath' phenomena, classified into Type I and II\footnote{\url{https://jwst-docs.stsci.edu/known-issues-with-jwst-data/nircam-known-issues/nircam-scattered-light-artifacts}}, resulting from off-field bright sources scattering light onto the detectors. We also observe the presence of ‘Ginkgo Leaf' artifacts, most commonly on detector A5, attributed to internal reflections from a nearby bright star, as well as ‘ghost stars' adjacent to bright stars. In addition, residual effects such as persistence, snowballs, wisps, and claws remain in some exposures, having evaded full correction during the initial reduction.  As the JWST pipeline, CRDS reference files, and our custom processing techniques continue to improve, complemented by systematic visual inspection, we expect to progressively suppress or eliminate these artifacts in subsequent data releases. Representative examples of these anomalies are presented in Fig.~\ref{Fig::artifacts}.

\section{Conclusions}
\label{sec:conclusion}

In this paper, we have outlined the assembly of the COSMOS-Web (\citealt{Casey2023}; PIs Kartaltepe \& Casey, ID=1727) mosaic utilizing JWST/NIRCam. Representing the largest continuous survey conducted with NIRCam to date, the COSMOS-Web survey adds a significant new layer of data to the already richly observed COSMOS field, benefiting from its extensive data coverage. The reduced data from the COSMOS-Web survey are publicly accessible via the following URL: \url{https://exchg.calet.org/cosmosweb-public/DR1/}.  The survey encompasses 152 pointings across four filters (F115W, F150W, F277W, F444W). The survey's layout is a 0.54 deg$^2$ rectangle arranged into 19 columns and 8 rows.

Data collection spanned three distinct epochs. The first ran from January 5, 2023 to January 6, 2023. The second period was from April 6 to April 23, 2023 (with an additional visit, CWEBTILE-0-4, on May 17, 2023), and the third epoch spanned from December 12, 2023 to January 7, 2024. The visits that were missed were observed afterwards from April 5, 2024 to May 17, 2024. The survey area was divided into 20 tiles, labeled from A1 to A10 and B1 to B10. We have made the processed images available in two resolutions, 30 mas and 60 mas.

This paper describes the data reduction techniques that were used to suit the specific needs of our survey. We also outline the characteristics of the COSMOS-Web survey, which achieves 5$\sigma$ depths of 26.7-28.2 for point sources within 0.15 arcsec apertures (without aperture correction).

The COSMOS-Web survey represents a major step forward in near- and mid-infrared extragalactic imaging, combining wide-area coverage with the depth and resolution of JWST. Thanks to its carefully designed observing strategy, the survey enables robust statistical studies of galaxy formation across cosmic time. As the largest JWST extragalactic program to date, its completion will provide a reference dataset for a broad range of follow-up studies.
\vspace{1cm}

We thank Sandro Tacchella and the NIRCam team for sharing the wisps templates used in the first version of the data reduction.  We acknowledge that the location where most of this work took place, the University of Texas at Austin, sits on the Indigenous lands of Turtle Island, the ancestral name for what now is called North America. Moreover, we would like to acknowledge the Alabama-Coushatta, Caddo, Carrizo/Comecrudo, Coahuiltecan, Comanche, Kickapoo, Lipan Apache, Tonkawa and Ysleta Del Sur Pueblo, and all the American Indian and Indigenous Peoples and communities who have been or have become a part of these lands and territories in Texas. This project has received funding from the European Union’s Horizon 2020 research and innovation program under the Marie Sklodowska-Curie grant agreement No 101148925. AR and DC acknowledge support from JWST-GO-06541 and JWST-GO-06585. DS carried out this research at the Jet Propulsion Laboratory, California Institute of Technology, under a contract with the National Aeronautics and Space Administration (80NM0018D0004). The French part of the COSMOS team is partly supported by the Centre National d'Etudes Spatiales (CNES).  This work was made possible by utilizing the CANDIDE cluster at the Institut d’Astrophysique de Paris, which was funded through grants from the PNCG, CNES, DIM-ACAV, and the Cosmic Dawn Center and maintained by S. Rouberol.


\appendix
\section{Astrometry per tiles}

\startlongtable
\begin{deluxetable*}{cccccccc}
\tabletypesize{\scriptsize}
\tablecaption{Astrometry\label{tab::astrometry_total}}
\tablewidth{0pt}
\tablehead{
\colhead{Tile}  & \colhead{Filter} & \colhead{median $\Delta$RA} &  \colhead{mean $\Delta$RA} &  \colhead{MAD RA} & \colhead{median $\Delta$Dec} &  \colhead{mean $\Delta$Dec} &  \colhead{MAD Dec} \\[-0.25cm]
 & & mas & mas & mas & mas & mas & mas 
}
\startdata
A1	&	F115W	&	2.24	&	1.04	&	11.32	&	0.12	&	2.11	&	11.13	\\
A1	&	F150W	&	1.70	&	1.09	&	11.70	&	0.55	&	1.23	&	11.86	\\
A1	&	F277W	&	0.30	&	0.21	&	12.07	&	0.09	&	0.20	&	12.31	\\
A1	&	F444W	&	0.15	&	-0.17	&	12.93	&	0.56	&	1.08	&	13.17	\\
A10	&	F115W	&	0.58	&	0.70	&	11.88	&	1.69	&	1.32	&	11.11	\\
A10	&	F150W	&	0.41	&	0.45	&	11.86	&	1.81	&	0.84	&	11.20	\\
A10	&	F277W	&	-0.86	&	0.08	&	12.16	&	1.38	&	0.77	&	11.68	\\
A10	&	F444W	&	0.37	&	1.06	&	13.34	&	0.11	&	-0.76	&	13.05	\\
A2	&	F115W	&	1.55	&	0.81	&	11.33	&	1.38	&	3.37	&	11.03	\\
A2	&	F150W	&	1.34	&	1.82	&	11.90	&	1.18	&	1.75	&	11.63	\\
A2	&	F277W	&	0.53	&	0.87	&	12.06	&	0.25	&	1.87	&	12.43	\\
A2	&	F444W	&	-0.14	&	0.63	&	12.92	&	-0.50	&	0.82	&	13.48	\\
A3	&	F115W	&	1.07	&	0.42	&	11.63	&	1.58	&	1.17	&	11.39	\\
A3	&	F150W	&	0.73	&	-0.44	&	11.38	&	1.39	&	0.74	&	11.85	\\
A3	&	F277W	&	-0.99	&	-2.59	&	12.25	&	0.51	&	1.06	&	12.20	\\
A3	&	F444W	&	-0.09	&	-1.53	&	13.52	&	0.24	&	0.74	&	13.38	\\
A4	&	F115W	&	1.42	&	1.82	&	10.78	&	1.20	&	1.50	&	10.70	\\
A4	&	F150W	&	1.25	&	1.45	&	11.12	&	1.01	&	0.85	&	11.33	\\
A4	&	F277W	&	0.39	&	-0.55	&	11.79	&	0.21	&	0.28	&	11.66	\\
A4	&	F444W	&	-0.84	&	-0.94	&	12.94	&	0.14	&	0.00	&	12.90	\\
A5	&	F115W	&	1.54	&	2.17	&	11.04	&	1.33	&	1.38	&	10.70	\\
A5	&	F150W	&	1.42	&	2.20	&	11.44	&	1.17	&	1.03	&	11.56	\\
A5	&	F277W	&	0.85	&	-0.21	&	11.74	&	0.69	&	0.86	&	11.75	\\
A5	&	F444W	&	0.65	&	0.15	&	13.16	&	0.26	&	-0.65	&	13.01	\\
A6	&	F115W	&	1.73	&	0.99	&	11.32	&	1.27	&	-0.36	&	11.45	\\
A6	&	F150W	&	1.61	&	1.84	&	11.66	&	1.50	&	-0.32	&	11.86	\\
A6	&	F277W	&	0.24	&	1.34	&	11.96	&	0.44	&	-0.65	&	12.33	\\
A6	&	F444W	&	0.04	&	1.07	&	13.25	&	0.66	&	-0.78	&	13.44	\\
A7	&	F115W	&	2.81	&	3.24	&	11.51	&	2.01	&	1.42	&	11.16	\\
A7	&	F150W	&	2.28	&	2.45	&	11.77	&	1.45	&	1.12	&	11.22	\\
A7	&	F277W	&	0.19	&	0.66	&	12.25	&	0.55	&	0.29	&	11.97	\\
A7	&	F444W	&	0.31	&	1.14	&	12.82	&	-0.50	&	-0.15	&	13.27	\\
A8	&	F115W	&	-0.38	&	-1.18	&	12.37	&	1.91	&	1.15	&	11.44	\\
A8	&	F150W	&	0.22	&	-0.65	&	12.41	&	2.20	&	1.30	&	11.96	\\
A8	&	F277W	&	-1.02	&	-1.64	&	12.64	&	0.02	&	-0.69	&	12.27	\\
A8	&	F444W	&	-0.90	&	-0.89	&	13.59	&	-0.42	&	-0.01	&	13.29	\\
A9	&	F115W	&	0.78	&	1.15	&	11.53	&	2.38	&	1.68	&	10.57	\\
A9	&	F150W	&	1.24	&	0.74	&	11.67	&	2.04	&	1.43	&	11.19	\\
A9	&	F277W	&	-0.42	&	-0.79	&	12.27	&	0.73	&	0.14	&	11.54	\\
A9	&	F444W	&	-1.03	&	-1.84	&	12.68	&	0.51	&	0.41	&	12.81	\\
B1	&	F115W	&	0.86	&	0.44	&	10.50	&	-1.95	&	-0.98	&	10.73	\\
B1	&	F150W	&	1.20	&	1.24	&	10.82	&	-1.69	&	-1.12	&	10.90	\\
B1	&	F277W	&	0.87	&	0.45	&	11.14	&	0.05	&	0.97	&	11.45	\\
B1	&	F444W	&	0.73	&	-0.28	&	12.41	&	-0.41	&	0.18	&	12.20	\\
B10	&	F115W	&	1.14	&	0.49	&	11.48	&	-2.15	&	-1.58	&	11.62	\\
B10	&	F150W	&	0.24	&	0.27	&	11.70	&	-1.37	&	-1.19	&	12.44	\\
B10	&	F277W	&	0.64	&	0.70	&	12.95	&	-0.16	&	-0.05	&	12.66	\\
B10	&	F444W	&	0.19	&	0.49	&	13.41	&	-0.36	&	-0.22	&	13.89	\\
B2	&	F115W	&	-1.89	&	-1.29	&	10.42	&	-1.65	&	-0.73	&	10.31	\\
B2	&	F150W	&	-1.98	&	-0.93	&	10.78	&	-2.09	&	-0.56	&	10.49	\\
B2	&	F277W	&	-0.32	&	0.13	&	10.73	&	-0.58	&	0.16	&	11.09	\\
B2	&	F444W	&	-0.17	&	-0.29	&	12.03	&	-1.44	&	-0.31	&	12.20	\\
B3	&	F115W	&	-0.47	&	-0.08	&	10.71	&	-2.80	&	-2.43	&	10.87	\\
B3	&	F150W	&	-0.03	&	0.70	&	11.19	&	-2.12	&	-0.74	&	11.53	\\
B3	&	F277W	&	0.78	&	1.58	&	11.60	&	-0.70	&	0.60	&	12.16	\\
B3	&	F444W	&	0.53	&	1.77	&	12.79	&	-0.65	&	1.20	&	13.19	\\
B4	&	F115W	&	0.48	&	0.65	&	11.36	&	-3.63	&	-3.41	&	11.10	\\
B4	&	F150W	&	0.29	&	0.22	&	11.82	&	-3.57	&	-3.28	&	11.44	\\
B4	&	F277W	&	-0.23	&	1.08	&	12.20	&	-1.39	&	-1.10	&	11.86	\\
B4	&	F444W	&	-0.03	&	-0.03	&	13.25	&	-0.97	&	-0.21	&	12.98	\\
B5	&	F115W	&	-1.67	&	-0.81	&	11.88	&	-1.54	&	-0.98	&	11.97	\\
B5	&	F150W	&	-1.37	&	-0.77	&	12.39	&	-1.39	&	-1.53	&	12.13	\\
B5	&	F277W	&	-0.82	&	-0.02	&	13.02	&	-0.29	&	0.98	&	12.93	\\
B5	&	F444W	&	0.33	&	0.18	&	13.78	&	-0.35	&	0.22	&	14.18	\\
B6	&	F115W	&	-0.61	&	0.48	&	10.23	&	-1.23	&	-2.06	&	10.54	\\
B6	&	F150W	&	-0.33	&	-0.62	&	11.06	&	-1.27	&	-1.78	&	11.46	\\
B6	&	F277W	&	0.13	&	0.14	&	11.56	&	-0.56	&	-0.90	&	11.79	\\
B6	&	F444W	&	-0.02	&	-0.83	&	12.63	&	-0.54	&	-0.98	&	12.92	\\
B7	&	F115W	&	-1.90	&	-1.94	&	10.69	&	-0.50	&	-0.47	&	10.66	\\
B7	&	F150W	&	-1.41	&	-1.27	&	11.19	&	-0.49	&	-0.18	&	11.77	\\
B7	&	F277W	&	-0.43	&	-1.05	&	11.14	&	0.17	&	-0.05	&	11.92	\\
B7	&	F444W	&	0.39	&	0.25	&	12.62	&	-0.17	&	0.33	&	13.10	\\
B8	&	F115W	&	-1.48	&	-2.40	&	10.71	&	-2.57	&	-2.98	&	11.33	\\
B8	&	F150W	&	-1.32	&	-1.89	&	11.44	&	-3.02	&	-3.46	&	12.09	\\
B8	&	F277W	&	-0.45	&	-1.06	&	12.01	&	-2.31	&	-3.05	&	12.07	\\
B8	&	F444W	&	0.09	&	0.58	&	13.14	&	-2.45	&	-2.94	&	13.40	\\
B9	&	F115W	&	0.71	&	-0.13	&	11.89	&	-2.49	&	-2.29	&	11.17	\\
B9	&	F150W	&	0.28	&	0.80	&	12.04	&	-3.13	&	-3.17	&	11.62	\\
B9	&	F277W	&	-0.17	&	0.00	&	12.47	&	-1.05	&	-1.29	&	12.14	\\
B9	&	F444W	&	-0.52	&	0.19	&	13.88	&	-0.52	&	-1.21	&	13.65	\\		
\enddata
\tablecomments{Overview of astrometric discrepancies identified between NIRCam observations in all four filters and the reference catalog across the 20 tiles of the COSMOS-Web survey, as elaborated in Section~\ref{subsec::astrometry}. Long Wavelength (LW) data were aligned directly with the reference catalog, while Short Wavelength (SW) data alignment utilized an intermediary catalog derived from F277W filter observations. This dual strategy ensured accurate astrometric calibration for both sets of observations within the survey.}
\end{deluxetable*}

\bibliography{cw_image}{}
\bibliographystyle{aasjournal}

\newpage

\allauthors

\end{document}